\documentclass{article}
\usepackage{arxiv}

\usepackage[utf8]{inputenc} 
\usepackage[T1]{fontenc}    
\usepackage{hyperref}       
\usepackage{url}            
\usepackage{datetime}       
\usepackage{booktabs}       
\usepackage{amsfonts}       
\usepackage{nicefrac}       
\usepackage{microtype}      
\usepackage{graphicx}
\usepackage{natbib}
\usepackage{doi}
\usepackage{xcolor}
\usepackage{bm}
\usepackage{cancel}
\usepackage{amssymb}

\usepackage{amsmath}
\usepackage{framed}
\usepackage[acronym]{glossaries}
\makeglossaries

\newacronym{fep}{FEP}{Free Energy Principle}
\newacronym{vfe}{VFE}{Variational Free Energy}
\newacronym{ness}{NESS}{Non-Equilibrium Steady-State}

\hypersetup{colorlinks = true,
linkcolor = blue,
urlcolor  = blue,
citecolor = blue,
anchorcolor = black}

\title{Self-orthogonalizing attractor neural networks emerging from the free energy principle}

\newdate{articleDate}{8}{3}{2026}
\date{\displaydate{articleDate}}

\makeatletter
\let\@fnsymbol\@arabic
\makeatother

\author{\href{https://orcid.org/0000-0002-2942-0821}{\includegraphics[scale=0.06]{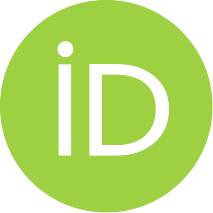}}\hspace{1mm}Tamas Spisak\footnotemark[1]\\
Center for Translational Neuro- and Behavioral Sciences (C-TNBS), University Medicine Essen, Germany\\\AND
\href{https://orcid.org/0000-0001-7984-8909}{\includegraphics[scale=0.06]{orcid.pdf}}\hspace{1mm}Karl Friston\\
Queen Square Institute of Neurology, University College London, WC1N 3AR, UK\\VERSES, Los Angeles, CA 90067, USA\\}

\renewcommand{\headeright}{}
\renewcommand{\undertitle}{}
\renewcommand{\shorttitle}{}

\hypersetup{
pdftitle={\@title},
pdfsubject={},
pdfauthor={\@author},
pdfkeywords={attractor networks,free energy principle,active inference,orthogonal representations,self-organization},
addtopdfcreator={Written in Curvenote}
}

\usepackage{caption}
\begin{document}
\maketitle
\footnotetext[1]{Correspondence to: tamas.spisak@uk-essen.de}

\begin{abstract}
Attractor dynamics are a hallmark of many complex systems, including the brain. Understanding how such self-organizing dynamics emerge from first principles is crucial for advancing our understanding of neuronal computations and the design of artificial intelligence systems. Here we formalize how attractor networks emerge from the free energy principle applied to a universal partitioning of random dynamical systems. Our approach obviates the need for explicitly imposed learning and inference rules and identifies emergent, but efficient and biologically plausible inference and learning dynamics for such self-organizing systems. These result in a collective, multi-level Bayesian active inference process. Attractors on the free energy landscape encode prior beliefs; inference integrates sensory data into posterior beliefs; and learning fine-tunes couplings to minimize long-term surprise. Analytically and via simulations, we establish that the proposed networks favor approximately orthogonalized attractor representations, a consequence of simultaneously optimizing predictive accuracy and model complexity. These attractors efficiently span the input subspace, enhancing generalization and the mutual information between hidden causes and observable effects. Furthermore, while random data presentation leads to symmetric and sparse couplings, sequential data fosters asymmetric couplings and non-equilibrium steady-state dynamics, offering a natural generalization of conventional Boltzmann Machines. Our findings offer a unifying theory of self-organizing attractor networks, providing novel insights for AI and neuroscience.
\end{abstract}


\begin{itemize}
\item Attractor networks are derived from the Free Energy Principle (\acrshort{fep}) applied to a universal partitioning of random dynamical systems.
\item This approach yields emergent, biologically plausible inference and learning dynamics, forming a multi-level Bayesian active inference process.
\item The networks favor approximately orthogonalized attractor representations, optimizing predictive accuracy and model complexity.
\item Sequential data presentation leads to asymmetric couplings and non-equilibrium steady-state dynamics, generalizing conventional Boltzmann Machines.
\item Simulations demonstrate orthogonal basis formation, generalization, sequence learning, scalability and resistance to catastrophic forgetting.
\end{itemize}

\textbf{Graphical Abstract}

\includegraphics[width=1.0\linewidth]{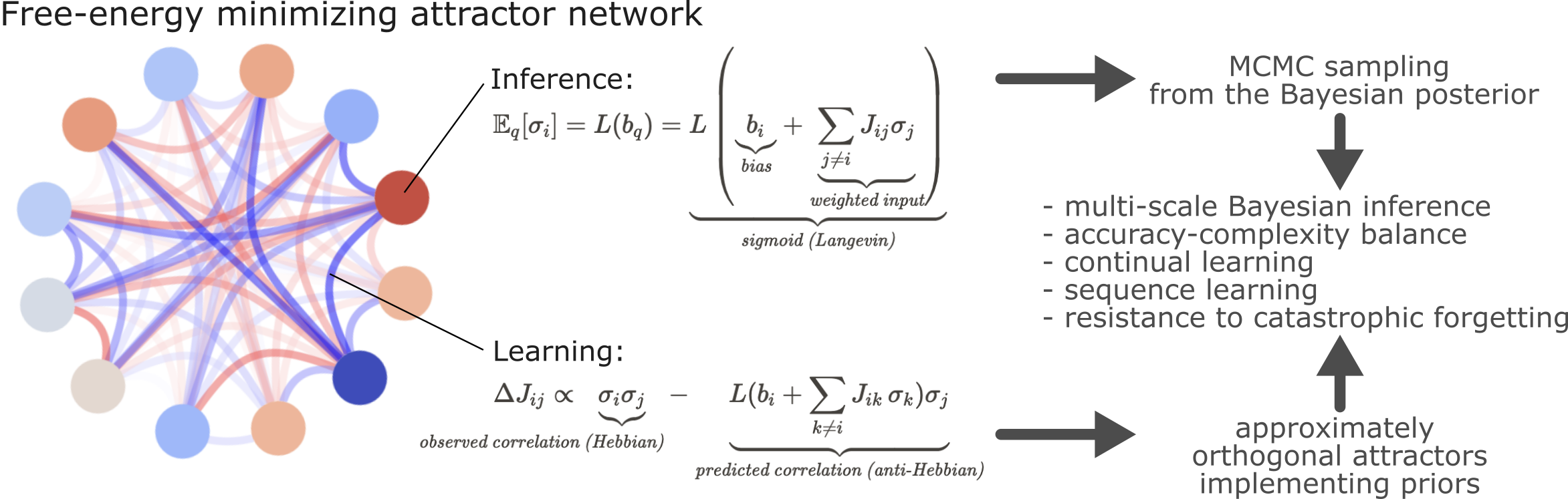}

\section{Introduction}

From whirlpools and bird flocks to neuronal and social networks, countless natural systems can be characterized by dynamics organized around \textit{attractor states} \citep{Haken_1978}. Such systems can be decomposed into a collection of - less or more complex - building blocks or ``particles'' (e.g. water molecules, birds, neurons, or people), which are coupled through local interactions. Attractors are an emergent consequence of the collective dynamics of the system, arising from these local interactions, without any individual particles exerting global control.

Attractors are a key concept in dynamical systems theory, defined as a set of states in the state space of the system to which nearby trajectories converge \citep{Guckenheimer_1984}. Geometrically, the simplest attractors are fixed points and limit cycles (representing periodic oscillations). However, the concept extends to more complex structures like strange attractors associated with chaotic behavior, as well as phenomena arising in stochastic or non-equilibrium settings, such as probability distributions over states (stochastic attractors), transient states reflecting past dynamics (ghost attractors or attractor ruins), and trajectories that cycle through sequences of unstable states (sequential attractors or heteroclinic cycles).
Artificial attractor neural networks \citep{Amit_1989} represent a class of recurrent neural networks specifically designed to leverage attractor dynamics. While the specific forms and behaviors of these networks are heavily influenced by the chosen inference and learning rules, self-organization is a key feature of all variants, as the stable states emerge from the interactions between network elements without explicit external coordination. This property makes them particularly relevant as models for self-organizing biological systems, including the brain.
It is clear that the brain is also a complex attractor network. Attractor dynamics have long been proposed to play a significant role in information integration at the circuit level \citep{Freeman_1987, Amit_1989, Deco_2003, https://doi.org/10.48550/arxiv.2504.12188} (\cite{Tsodyks_1999}) and have become established models for canonical brain circuits involved in motor control, sensory amplification, motion integration, evidence integration, memory, decision-making, and spatial navigation (see \citet{Khona_2022} for a review). For instance, the activity of head direction cells - neurons that fire in a direction-dependent manner - are known to arise from a circular attractor state, produced by a so-called ring attractor network \citep{Zhang_1996}. Multi- and metastable attractor dynamics have also been proposed to extend to the meso- and macro-scales \citep{Rolls_2009}, ``accommodating the coordination of heterogeneous elements'' \citep{Kelso_2012}, rendering attractor dynamics an overarching computational mechanism across different scales of brain function.
The brain, as an instance of complex attractor networks, showcases not only the computational capabilities of this network architecture but also its ability to emerge and evolve through self-organization.

When discussing self-organization in attractor networks, we will differentiate two distinct levels. First, we can talk about \textit{operational self-organization}: the capacity of a pre-formed network to settle into attractor states during its operation. This however does not encompass the network's ability to ``build itself'' -- to emerge from simple, local interactions without explicit programming or global control, and to adaptively evolve its structure and function through learning. This latter level of self-organization is what we will refer to as \textit{adaptive self-organization}.
Architectures capable of adaptive self-organization would mirror the nervous system's capacity to not just function as an attractor network, but to become - and remain - one through a self-directed process of development and learning. Further, adaptive self-organization would also be a highly desirable property for robotics and artificial intelligence systems, not only boosting their robustness and adaptability by means of continuous learning, but potentially leading to systems that can increase their complexity and capabilities organically over time (e.g. developmental robotics).
Therefore, characterizing \textit{adaptive self-organization} in attractor networks is vital for advancing our understanding of the brain and for creating more autonomous, adaptive, brain-inspired AI systems.

The Free Energy Principle (\acrshort{fep}) offers a general framework to study self-organization to non-equilibrium steady states as Bayesian inference (a.k.a., active inference). The \acrshort{fep} has been pivotal in connecting the dynamics of complex self-organizing systems with computational and inferential processes, especially within the realms of brain function \citep{Friston_2023, Friston_2012, Palacios_2020}. The \acrshort{fep} posits that any `thing' - in order to exist for an extended period of time - must maintain conditional independence from its environment. This entails a specific sparsely coupled structure of the system, referred to as a \textit{particular partition}, that divides the system into internal, external, and boundary (sensory and active) states (see Figure~\ref{fig-concept}A). It can be shown that maintaining this sparse coupling is equivalent to executing an inference process, where internal states deduce the causes of sensory inputs by minimizing variational free energy (see \citet{Friston_2023} or \citep{Friston_2023} for a formal treatment).

\begin{figure}[!htbp]
\centering
\includegraphics[width=0.9\linewidth]{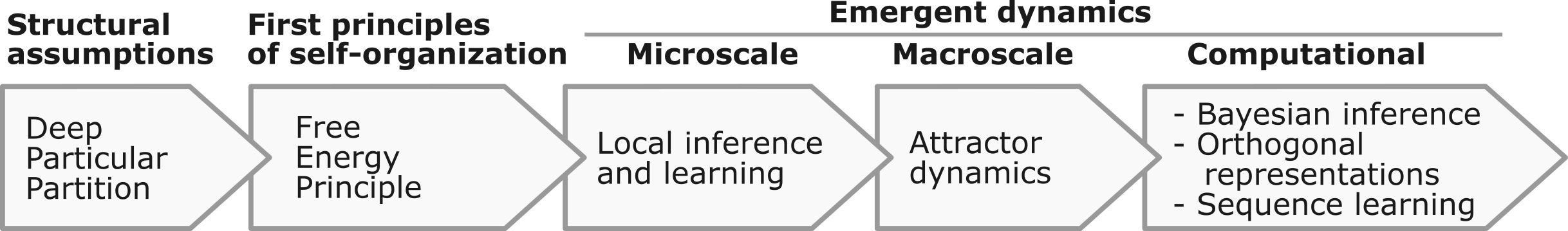}
\caption[]{\textbf{Overview of the framework.} Starting from universal and parsimonious structural assumptions (deep particular partition), variational free energy minimization under the Free Energy Principle (\acrshort{fep}) gives rise to emergent dynamics at multiple scales: local stochastic inference and learning rules that guide the update of internal states and the couplings between them, self-organizing attractor dynamics at the macroscale, and --- at the computational level --- Bayesian inference, approximately orthogonal attractor representations of latent external causes, and sequence learning capabilities.}
\label{fig-framework-overview}
\end{figure}

Here, we describe the specific class of adaptive self-organizing attractor networks that emerge directly from the \acrshort{fep}, without the need for explicitly imposed learning or inference rules (Figure~\ref{fig-framework-overview}).
First, we show that a hierarchical formulation of \textit{particular partitions} - a concept that is applicable to any complex dynamical system - can give rise to systems that have the same functional form as well-known artificial attractor network architectures.
Second, we show that minimizing variational free energy (\acrshort{vfe}) with regard to the internal states of such systems yields a Boltzmann Machine-like stochastic update mechanism, with continuous-state stochastic Hopfield networks being a special case.
Third, we show that minimizing \acrshort{vfe} with regard to the internal blanket or boundary states of the system (couplings) induces a generalized predictive coding-based learning process. Crucially, this adaptive process extends beyond simply reinforcing concrete sensory patterns; it learns to span the entire subspace of key patterns by establishing approximately-orthogonalized attractor representations, which the system can then combine during inference.
We use simulations to identify the requirements for the emergence of quasi-orthogonal attractors and to illustrate the derived attractor networks' ability to generalize to unseen data. Finally, we highlight that the derived attractor network can naturally produce sequence-attractors (if the input data is presented in a clear sequential order) and exemplifies its potential to perform continual learning and to overcome catastrophic forgetting by means of spontaneous activity.
We conclude by discussing testable predictions of our framework, and exploring the broader implications of these findings for both natural and artificial intelligence systems.

\section{Main Results}

\subsection{Background: particular partitions and the free energy principle}

\begin{figure}[!htbp]
\centering
\includegraphics[width=0.66\linewidth]{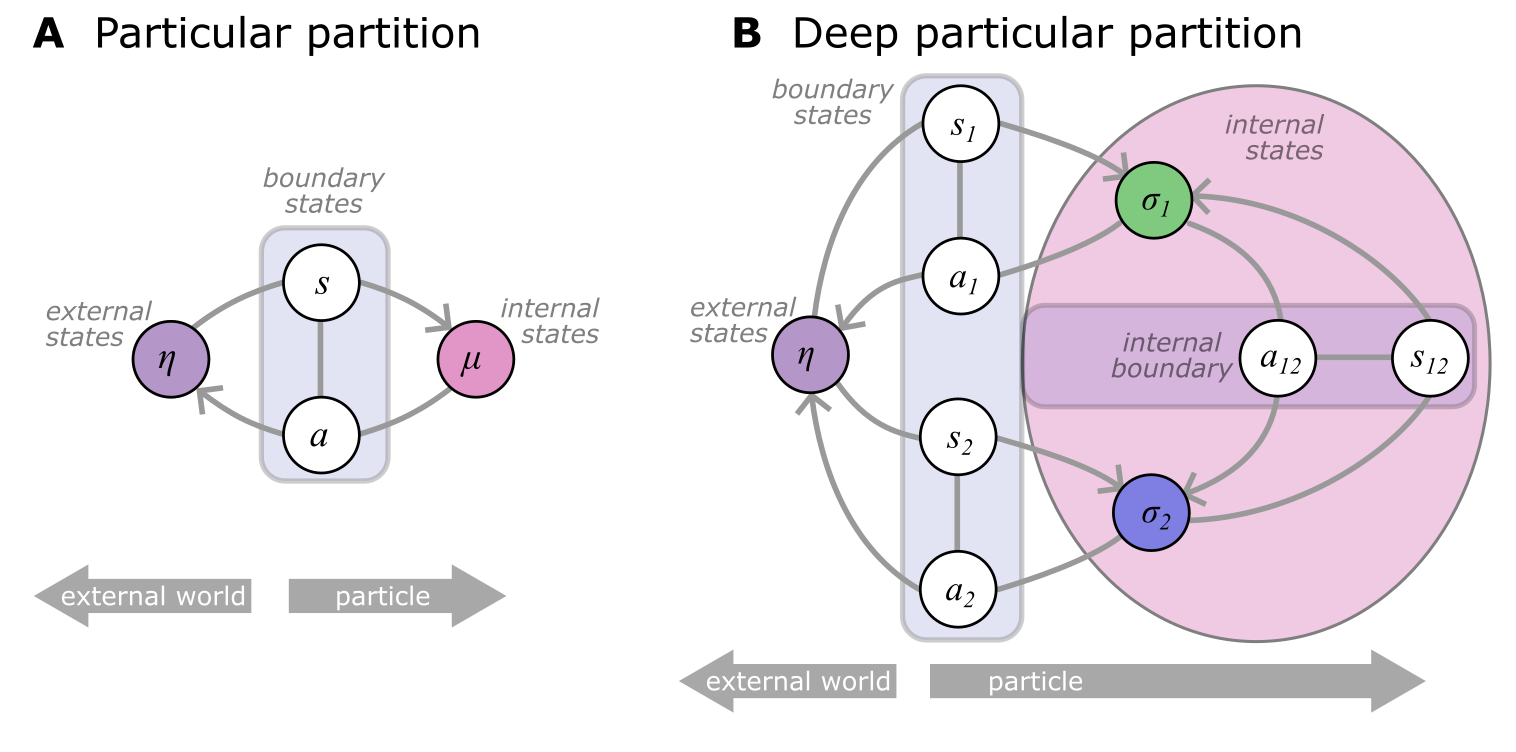}
\caption[]{\textbf{Deep Particular Partitions.} \newline
\textbf{A} Schematic illustration of a particular partition of a system into internal ($\mu$) and external states ($\eta$), separated by a Markov blanket consisting of sensory states ($s$) and active states ($a$). The tuple $(\mu, s, a)$ is called a \textit{particle}. A particle, in order to persist for an extended period of time, will necessarily have to maintain its Markov blanket, a behavior that is equivalent to an inference process in which internal states infer external states through the blanket states. The resulting self-organization of internal states corresponds to perception, while actions link the internal states back to the external states.
\textbf{B} The internal states $\mu \subset x$ can be arbitrarily complex. Without loss of generality, we can consider that the macro-scale $\mu$ can be decomposed into set of overlapping micro-scale \textit{subparticles} ($\sigma_i, s_i, a_i, s_{ij}, a_{ij}$), so that the internal state of subparticle $\sigma_i \subset \mu$ can be an external state from the perspective of another subparticle $\sigma_j \subset \mu$. Some, or all subparticles can be connected to the macro-scale external state $\eta$, through the macro-scale Markov blanket, giving a decomposition of the original boundary states into $s_i \subset s$ and $a_i \subset a$. The subparticles are connected to each other by the micro-scale boundary states $s_{ij}$ and $a_{ij}$. Note that this notation considers the point-of-view of the $i$-th subparticle. Taking the perspective of the $j$-th subparticle, we can see that $s_{ji}=a_{ij}$ and $a_{ji}=s_{ij}$. While the figure depicts the simplest case of two nested partitions, the same scheme can be applied recursively to any number of (possibly nested) subparticles and any coupling structure amongst them.}
\label{fig-concept}
\end{figure}

Our effort to characterize self-organizing attractor networks calls for an individuation of `self' from nonself. \textit{Particular partitions}, a concept that is at the core of the Free Energy Principle (\acrshort{fep}) \citep{https://doi.org/10.48550/arxiv.2210.12761, Friston_2023, Friston_2012, Palacios_2020}, is a natural way to formalize this individuation.
As a toy intuition, one can think of a recurrent neural assembly (internal states) interacting with incoming sensory drives and outgoing action channels (blanket states), while latent external causes remain hidden. The blanket states mediate all interactions, which is exactly the structure needed for inference to be well-defined.
A particular partition is a partition that divides the states of a system $x$ into a particle or `thing' $(s,a,\mu) \subset x$ and its external states $\eta \subset x$, based on their sparse coupling:

\begin{align*}
\begin{bmatrix}
\dot{\eta}(\tau) \\
\dot{s}(\tau) \\
\dot{a}(\tau) \\
\dot{\mu}(\tau)
\end{bmatrix}
=
\begin{bmatrix}
f_{\eta}(\eta, s, a) \\
f_{s}(\eta, s, a) \\
f_{a}(s, a, \mu) \\
f_{\mu}(s, a, \mu)
\end{bmatrix}
+
\begin{bmatrix}
\omega_{\eta}(\tau) \\
\omega_{s}(\tau) \\
\omega_{a}(\tau) \\
\omega_{\mu}(\tau)
\end{bmatrix}
\end{align*}

where $\eta$, $\mu$, $s$, and $a$ are the external, internal, sensory, and active states of the particle, respectively. The fluctuations $\omega_i, i \in (\mu, s, a, \eta)$ are assumed to be mutually independent. The particular states $\mu$, $s$, and $a$ are coupled to each other with \textit{particular flow dependencies}; namely, external states can only influence themselves and sensory states, while internal states can only influence themselves and active states (see Figure~\ref{fig-concept}A). It can be shown that these coupling constraints mean that external and internal paths are statistically independent, when conditioned on $s$ and $a$, often referred to as the Markov blanket \citep{https://doi.org/10.48550/arxiv.2210.12761}:

\begin{equation}
\label{eq-conditional-independence}
\eta \perp \mu \mid s, a.
\end{equation}

At an intuitive level, considering particular partitioning as a ``universal partitioning'' means that any persistent random dynamical system can be described with (possibly nested) interfaces that separate what is inferred from what is inferred about. In this view, Markov blankets are not ad hoc modeling devices, but the minimal bookkeeping needed to explain how local interactions can support stable, self-organizing inference across scales.

As shown by \citet{Friston_2023}, such a particle, in order to persist for an extended period of time, will necessarily have to maintain this conditional independence structure, a behavior that is equivalent to an inference process in which internal states infer external states through the blanket states (i.e., sensory and active states) by minimizing free energy \citep{Friston_2009, Friston_2010, Friston_2023}:

\begin{equation}
\label{eq-free-energy-principle}
\eta \perp \mu \mid s, a \quad \Rightarrow \quad \dot{\mu} = -\nabla_{\mu} F(s, a, \mu)
\end{equation}

where $F(s, a, \mu)$ is the variational free energy (\acrshort{vfe}):

\begin{equation}
\label{eq-free-energy-functional}
F(s,a,\mu) = \mathbb{E}_{q_\mu(\eta)}[\ln q_\mu(\eta) - \ln p(s,a,\eta)]
\end{equation}

with $q_\mu(\eta)$ being a variational density over the external states that is parameterized by the internal states and $p(s,a,\eta)$ being the joint probability distribution of the sensory, active and external states, a.k.a. the generative model \citep{Friston_2016}.

\subsection{Deep particular partitions and subparticles}

Particular partitions provide a universal description of complex systems, in the sense that the internal states $\mu$ behave as if they are inferring the external states under a generative model; i.e., a `black box' inference process (or computation), which can be arbitrarily complex. At the same time, the concept of particular partitions speaks to a recursive composition of ensembles (of things) at increasingly higher spatiotemporal scales \citep{https://doi.org/10.48550/arxiv.1906.10184, Palacios_2020, https://doi.org/10.15502/9783958573031}, which yields a natural way to resolve the internal complexity of $\mu$. Partitioning the ``macro-scale'' particle $\mu$ into multiple, overlapping ``micro-scale'' \textit{subparticles} $\{\pi_i\}_{i=1}^n$ - that themselves are particular partitions - we can unfold the complexity of the macro-scale particle to an arbitrary degree. As subparticles can be nested arbitrarily deep - yielding a hierarchical generative model - we refer to such a partitioning as a \textit{deep particular partition}.

As illustrated in Figure~\ref{fig-concept}B, each subparticle $\pi_i$ has internal states $\sigma_i$, and the coupling between any two subparticles $i$ and $j$ is mediated by micro-scale boundary states: sensory states $s_{ij}$ (representing the sensory information in $i$ coming from $j$) and active states $a_{ij}$ (representing the action of $i$ on $j$). The boundary states of subparticles naturally overlap; the sensory state of a subparticle $\sigma_i$ is the active state of $\sigma_j$ and vice versa, i.e. $a_{ji}=s_{ij}$ and $s_{ji}=a_{ij}$. This also means that, at the micro-scale, the internal state of a subparticle $\sigma_i \subset \mu$ is part of the external states for another subparticle in $\sigma_j \subset \mu$. Accordingly, the internal state of a subparticle $\sigma_i$ is conditionally independent of any other internal states $\sigma_j$ with $j \neq i$, given the blanket states of the subparticles:

\begin{equation}
\sigma_i \perp \sigma_j \mid a_{ij}, s_{ij}
\end{equation}

Note that this definition embraces sparse couplings across subparticles, as $a_{ij}$ and $s_{ij}$ may be empty for a given $j$ (no direct connection between the two subparticles), but we require the subparticles to yield a \textit{complete coverage} of $\mu$: $\bigcup_{i=1}^n \pi_i = \mu$.

\subsection{The emergence of attractor neural networks from deep particular partitions}

\begin{figure}[!htbp]
\centering
\includegraphics[width=0.5\linewidth]{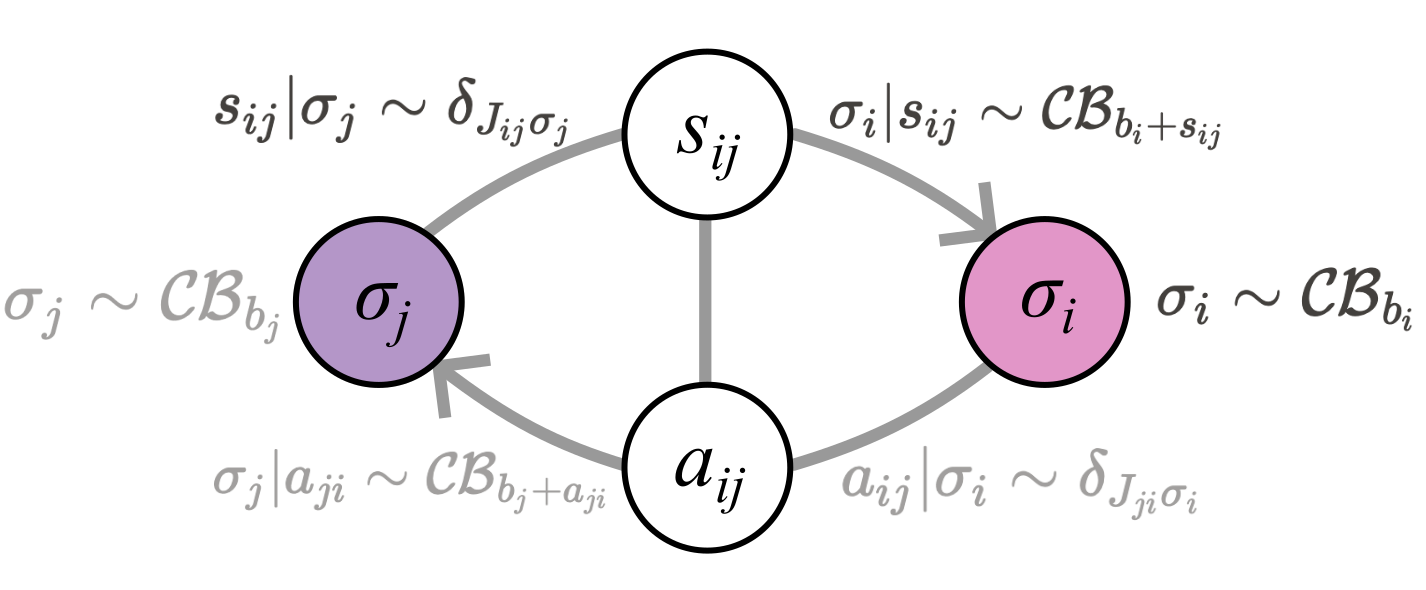}
\caption[]{\textbf{Parametrization of subparticles in a deep particular partition.} \newline
The internal state $\sigma_i$ of subparticle $\pi_i$ follows a continuous Bernoulli distribution, (a.k.a.\ a truncated exponential distribution supported on the interval $[-1,+1] \subset \mathbb{R}$, see Appendix-1, with a prior ``bias'' $b_i$ that can be interpreted as a priori log-odds evidence for an event (stemming from a macro-scale sensory input $s_{i}$---not shown, or from the internal dynamics of $\sigma_i$ itself, e.g.\ internal sequence dynamics).
The state $\sigma_i$ is coupled to the internal state of another subparticle $\sigma_j$ through the micro-scale boundary states $s_{ij}$ and $a_{ij}$. The boundary states simply apply a deterministic scaling to their respective $\sigma$ state, with a weight ($J_{ij}$) implemented by a Dirac delta function shifted by $J_{ij}$. The state $\sigma_i$ is influenced by its sensory input $s_i$ in a way that $s_i$ gets integrated into its internal bias, updating the level of evidence for the represented event.}
\label{fig-parametrization}
\end{figure}

Next, we establish a prototypical mathematical parametrization for an arbitrary deep particular partition, shown on Figure~\ref{fig-parametrization}, with the aim of demonstrating that such complex, sparsely coupled random dynamical systems can give rise to artificial attractor neural networks.

In our example parametrization, we assume that the internal states of subparticles in a complex particular partition are \textit{continuous} Bernoulli states (also known as a truncated exponential distribution, see \href{https://pni-lab.github.io/fep-attractor-network/appendix/#appendix-1}{Appendix~1} for details), denoted by $\sigma_i \sim \mathcal{CB}_{b_i}$:

\begin{equation}
\label{prior-sigma}
p(\sigma_i) \propto e^{b\sigma_i}
\end{equation}

Here, $\sigma_i \in [ -1, +1] \subset \mathbb{R}$, and $b_i \in \mathbb{R}$ represents an a priori bias (e.g. the level of prior log-odds evidence for an event) in $\sigma_i$ Figure~\ref{fig-parametrization}. Zero bias represents a flat prior (uniform distribution over [-1,1]), while positive or negative biases represent evidence for or against an event, respectively.
The above probability is defined up to a normalization constant, $b_i / (2 \sinh(b_i))$. See \href{https://pni-lab.github.io/fep-attractor-network/appendix/#appendix-2}{Appendix~2} for details.

Note that the assumption of continuous Bernoulli distribution is very parsimonious; it emerges directly from \acrshort{fep} through as the constrained maximum entropy solution to a single constraint on mean, plus finite support \citep{Ramstead_2023}.

Next, we define the conditional probabilities of the sensory and active states, creating the boundary between two subparticles $\sigma_i$ and $\sigma_j$: $s_{ij}|\sigma_j \sim \mathcal{\delta}_{J_{ij}\sigma_j}$ and $a_{ij} |\sigma_i \sim \mathcal{\delta}_{J_{ji}\sigma_i}$, where $\mathcal{\delta}$ is the Dirac delta function and $\bm{J}$ is a weight matrix. The elements $J_{ij}$ contains the weights of the coupling matrix (see \href{https://pni-lab.github.io/fep-attractor-network/appendix/#appendix-3}{Appendix~3}).

Expressed as PDFs:

\begin{equation}
p(s_{ij} | \sigma_j) = \delta(s_{ij} - J_{ij}\sigma_j)
\end{equation}
\begin{equation}
p(a_{ij} | \sigma_i) = \delta(a_{ij} - J_{ji}\sigma_i)
\end{equation}

The choice of this deterministic parametrization means that we introduce the assumption of the subparticles being conservative particles, as defined in \citet{Friston_2023}.

To close the loop, we define how the internal state $\sigma_i$ depends on its sensory input $s_{ij}$. We assume the sensory input simply adds to the prior bias $\sigma_i | s_{ij} \sim \mathcal{CB}_{b_i + s_{ij}}$:

\begin{equation}
p(\sigma_i | s_{ij}) \propto e^{(b_i + s_{ij})\sigma_i}
\end{equation}

With the continuous Bernoulli distribution, this simply means that the sensory evidence $s_{ij}$ adds to (or subtracts from) the prior belief $b_i$.

We now write up the direct conditional probability describing $\sigma_i$ given $\sigma_j$, marginalizing out the sensory and active states:

\begin{align}
\label{sigma-given-mu}
p(\sigma_i | \sigma_j) &= \int p(\sigma_i | s_{ij}) p(s_{ij} | \sigma_j) \, d s_{ij} \\
&\propto \int e^{(s_{ij}+b_i)\sigma_i} \delta( s_{ij} - J_{ij} \sigma_j ) d s_{ij} \\
&\propto e^{(b_i + J_{ij}\sigma_j)\sigma_i}
\end{align}

\begin{framed}
\textbf{Note}\\
The expected value of $p(\sigma_i | \sigma_j)$ is a sigmoid function of $\sigma_j$ \href{https://pni-lab.github.io/fep-attractor-network/appendix/#appendix-4}{Appendix~4}, specifically the Langevin function. This property allows it to function as an activation function in neural networks, enabling the network to model more complex patterns and decision boundaries.
\end{framed}

More generally, when sub-particle $i$ receives sensory input from all $n -1$ other sub-particles, the combined evidence at node $i$ is:

\begin{equation}
\label{total-evidence}
\theta_i = b_i + \sum_{j \neq i} J_{ij} \sigma_j
\end{equation}

The local conditional $p(\sigma_i | \sigma_{\backslash i}) \propto \exp(\theta_i \sigma_i)$ is a continuous Bernoulli with parameter $\theta_i$, and its expected value is $\mathbb{E}[\sigma_i] = L(\theta_i)$, the Langevin function. This local evidence form is the basis for the inference and learning rules derived in the following sections: they follow from \acrshort{vfe} minimization at each node using only $\theta_i$, and depend on the full (potentially asymmetric) coupling matrix $\bm{J}$. The global consequences of the coupling structure --- including the stationary distribution and non-equilibrium steady-state dynamics --- are discussed in section \ref{self-solenoidization}.

\subsection{Inference}

So far our derivation relied only on the sparsely coupled structure of the system (deep particular partition), but did not utilize the free energy principle itself. We now invoke the \acrshort{fep} to derive how each node updates its state. From the perspective of node $\sigma_i$, the local variational free energy is (see eq. (\ref{eq-free-energy-functional})):

\begin{equation}
\label{free-energy-rnn}
F_i[q_i] = \mathbb{E}_{q_i}\!\left[-\log p(\sigma_i \mid \theta_i)\right] - H[q_i]
\end{equation}

where $q_i$ is the variational density over $\sigma_i$, $H[q_i]$ its entropy, and $\theta_i = b_i + \sum_{j \neq i} J_{ij} \sigma_j$ is the total evidence at node $i$ (eq. (\ref{total-evidence})). For the continuous Bernoulli likelihood, $-\log p(\sigma_i \mid \theta_i) = \psi(\theta_i) - \theta_i \sigma_i$ (up to a constant, where $\psi(\theta) = \log \frac{2\sinh\theta}{\theta}$ is the log-partition function), giving:

\begin{equation}
F_i[q_i] = \psi(\theta_i) - \theta_i\,\mathbb{E}_{q_i}[\sigma_i] - H[q_i]
\end{equation}

Minimizing $F_i$ over all normalized densities $q_i$ supported on $[ -1,1]$ yields (see \href{https://pni-lab.github.io/fep-attractor-network/appendix/#appendix-6}{Appendix~6} for the full derivation via accuracy--complexity decomposition):

\begin{equation}
q_i^*(\sigma_i) \propto \exp(\theta_i \sigma_i)\,\mathbf{1}_{[-1,1]}(\sigma_i) = p(\sigma_i \mid \theta_i)
\end{equation}

That is, the optimal variational density is the continuous Bernoulli with parameter $\theta_i$, recovering the local posterior family that can be derived from the constrained maximum entropy construction \citep{Ramstead_2023}. The expected value under this optimal posterior gives the inference rule:

\begin{equation}
\label{inference-rule}
\boxed{
\mathbb{E}_{q}[\sigma_i] = L(\theta_i) = \underbrace{ L \left( \underbrace{ b_i}_{\textit{bias}} + \underbrace{\sum_{j \ne i} J_{ij} \sigma_j}_{\textit{weighted input}} \right) }_{ \textit{sigmoid (Langevin)} } 
}
\end{equation}

This derivation is purely local: it uses only the conditional $p(\sigma_i \mid \theta_i)$ at each node and does not require the global joint distribution (eq. (\ref{hopfield-joint})). In the case of symmetric couplings, it reduces to the familiar Boltzmann-style update rule of a continuous-state stochastic Hopfield network. The Langevin sigmoid arises naturally from the continuous Bernoulli parameterization, extending the standard deterministic gradient descent on the energy landscape to a full probabilistic framework (see \href{https://pni-lab.github.io/fep-attractor-network/appendix/#appendix-6}{Appendix~6}).
The resulting stochastic dynamics represents a local approximate Bayesian inference, where each node balances prior information ($b_i$) with evidence from neighboring nodes ($\sum J_{ij} \sigma_j$). As we show below, this inherent stochasticity allows the network to escape local energy minima --- consistent with MCMC methods --- thereby enabling macro-scale Bayesian inference.

\subsection{Learning}

At optimum, $q$ would match $p$, causing the \acrshort{vfe}'s derivative to vanish.
Learning happens, when there is a systematic change in the prior bias $b_i$ that counteracts the update process. This can correspond, for instance, to an external input (e.g. sensory signal representing increased evidence for an external event), but also as the result of the possibly complex internal dynamics of a subparticle (e.g. internal sequence dynamics or memory retrieval). In this case, a subparticle can take use of another (slower) process, to decrease local \acrshort{vfe}: it can change the way its action states are generated; and rely on its vicarious effects on sensory signals. In our parametrization, this can be achieved by changing the coupling strength $J_{ji}$ corresponding to the action states. Of note, while changing $J_{ji}$ corresponds to a change in action-generation at the local level of the subparticle (micro-scale), at the macro-scale, it can be considered as a change in the whole system's generative model.

Using the local \acrshort{vfe} from eq. (\ref{free-energy-rnn}) with a point-mass (sample-based) estimate for $\sigma_i$, the reduced objective for the coupling gradient is:

\begin{equation}
F_i(\theta_i; \sigma_i) = \psi(\theta_i) - \theta_i \sigma_i
\end{equation}

where $\theta_i = b_i + \sum_{j \neq i} J_{ij} \sigma_j$ is the total evidence at node $i$. A perturbation $\delta J_{ij}$ produces $\delta \theta_i = \sigma_j \delta J_{ij}$, and by the chain rule:

\begin{equation}
\frac{dF_i}{dJ_{ij}} = \frac{\partial F_i}{\partial \theta_i}\frac{\partial \theta_i}{\partial J_{ij}} = \bigl[L(\theta_i)-\sigma_i\bigr]\,\sigma_j
\end{equation}

Gradient descent on $F_i$ yields:

\begin{equation}
\label{learning-rule}
\boxed{
\Delta J_{ij} \propto \underbrace{\sigma_i \sigma_j}_{\textit{observed correlation (Hebbian)}} - \underbrace{ L(b_i + \sum_{k\neq i} J_{ik}\,\sigma_k ) \sigma_j}_{\textit{predicted correlation (anti-Hebbian)}}
}
\end{equation}

This learning resembles the family of ``Hebbian / anti-Hebbian'' or ``contrastive'' learning rules and it explicitly implements predictive coding (akin to \textit{prospective configuration} \citep{Song_2024, https://doi.org/10.48550/arxiv.2207.12316}). However, as opposed to e.g. contrastive divergence (a common method for training certain types of Boltzmann machines, \citet{Hinton_2002}), it does not require to contrast longer averages of separate ``clamped'' (fixed inputs) and ``free'' (free running) phases, but rather uses the instantaneous correlation between presynaptic and postsynaptic activation to update the weight, lending a high degree of scalability for this architecture. As we demonstrate below with \href{https://pni-lab.github.io/fep-attractor-network/simulation-demo}{Simulation 1}-4, this learning rule converges to symmetric weights (akin to a classical stochastic continuous-state Hopfield network), if input data is presented in random order and long epochs. At the same time, if data is presented in rapidly changing fixed sequences (\href{https://pni-lab.github.io/fep-attractor-network/simulation-digits-continuous-sequence}{Simulation 3}), the learning rule results in temporal predictive coding and learns asymmetric weights, akin to \citep{Millidge_2024}. As discussed above, in this case the symmetric component of $J$ encodes fixed-point attractors and the probability flux induced by the antisymmetric component results in sequence dynamics (conservative solenoidal flow), without altering the steady state of the system.

Other key features of this rule are: (i) its resemblance to Sanger's rule \citep{Sanger_1989} - an online algorithm for principal component analysis - and (ii) the analogy of its anti-Hebbian component to unlearning (``remotion'') processes in ``dreaming attractor networks'' \citep{Hopfield_1983, Plakhov, Dotsenko_1991, Fachechi_2019}, which serve to eliminate spurious system attractors. Such networks can effectively approximate Kanter-Sompolinsky projector neural networks \citep{Personnaz_1985, Kanter_1987}, characterized by orthogonal attractor states and maximal memory capacity. These parallels suggest that the derived free energy minimizing learning rule will promote the development of approximately orthogonal attractor states. We motivate this theoretically and demonstrate it with simulations in the subsequent sections.
In biological terms, the first term of eq. (\ref{learning-rule}) corresponds to correlation-dependent potentiation, while the second term implements a decorrelating, predictive-error-like correction that counterbalances runaway reinforcement. This places the rule in the algorithmic family of Hebbian/anti-Hebbian and homeostatically stabilized plasticity motifs, while we do not claim a one-to-one mapping to any single microcircuit mechanism. In other words, the derivation is normative (from free-energy minimization), and biological plausibility is asserted at the level of local computable updates and observed qualitative motifs.

\subsection{Emergence of approximately orthogonal attractors}

Under the \acrshort{fep}, learning not only aims to maximize accuracy but also minimizes complexity (\ref{free-energy-rnn}), leading to parsimonious internal generative models (encoded in the weights $\mathbf{J}$ and biases $\mathbf{b}$) that offer efficient representations of environmental regularities. A generative model is considered more complex (and less efficient) if its internal representations, specifically its attractor states $\boldsymbol{\sigma}^{(\mu)}$ - which loosely correspond to the modes of $p(\boldsymbol{\sigma})$ - are highly correlated or statistically dependent. Such correlations imply redundancy, as distinct attractors would not be encoding unique, independent features of the input space. Our learning rule (eq. (\ref{learning-rule})) - by minimizing micro-scale \acrshort{vfe} - inherently also minimizes the complexity term $D_{KL}[q(\sigma_i) || p(\sigma_i)]$, which regularizes the inferred state $q(\sigma_i)$ to be close to the node's current prior $p(\sigma_i)$. This not only induces sparsity in the coupling matrix, but - as we motivate below - also penalizes overlapping attractor representations and favours orthogonal representations. Hebbian learning - in itself - can not implement such a behavior; as it simply aims to store the current pattern by strengthening connections between co-active nodes. This has to be counter-balanced by the anti-Hebbian term, that subtracts the variance that is already explained out by the network's predictions.

To illustrate how this dynamic gives rise to efficient, (approximately) orthogonal representations of the external states, suppose the network has already learned a pattern $\mathbf{s}^{(1)}$, whose neural representation is the attractor $\boldsymbol{\sigma}^{(1)}$ and associated weights $\mathbf{J}^{(1)}$. When a new pattern $\mathbf{s}^{(2)}$ is presented that is correlated with $\mathbf{s}^{(1)}$, the network's prediction for $\sigma_i^{(2)}$ will be $\hat{\sigma}_i = L(b_i + \sum_{k \neq i} J_{ik}^{(1)} \sigma_k)$. Because inference with $\mathbf{J}^{(1)}$ converges to $\boldsymbol{\sigma}^{(1)}$ and $\boldsymbol{\sigma}^{(2)}$ is correlated with $\boldsymbol{\sigma}^{(1)}$, the prediction $\hat{\boldsymbol{\sigma}}$ will capture variance in $\boldsymbol{\sigma}^{(2)}$ that is `explained' by $\boldsymbol{\sigma}^{(1)}$.
The learning rule updates the weights based only on the unexplained (residual) component of the variance, the prediction error. In other words, $\hat{\boldsymbol{\sigma}}$ approximates the projection of $\boldsymbol{\sigma}^{(2)}$ onto the subspace already spanned by $\boldsymbol{\sigma}^{(1)}$. Therefore, the weight update primarily strengthens weights corresponding to the component of $\boldsymbol{\sigma}^{(2)}$ that is orthogonal to $\boldsymbol{\sigma}^{(1)}$.
Thus, the learning effectively encodes this residual, $\boldsymbol{\sigma}^{(2)}_{\perp}$, ensuring that the new attractor components being formed tend to be orthogonal to those already established.
As we demonstrate in the next section with \href{https://pni-lab.github.io/fep-attractor-network/simulation-demo}{Simulation 1}-2, repeated application of this rule during learning progressively decorrelates the neural activities associated with different patterns, leading to \textbf{approximately orthogonal attractor states}. This process is analogous to online orthogonalization procedures (e.g., Sanger's rule for PCA \citep{Sanger_1989}), as well as the converge of \citet{Plakhov}'s unlearning rule to the projector (or pseudo-inverse) matrix of stored patterns \citep{Fachechi_2019}. As a result, free energy minimizing attractor networks yield a powerful stochastic approximation of the ``projector networks'' of \citet{Personnaz_1985}, \citet{Kanter_1987}, which offer maximal memory capacity and retrieval without errors.

While orthogonality enhances representational efficiency, it raises the question of how the network retrieves the original patterns from these (approximately) orthogonalized representations - a key requirement to function as associative memory. As discussed next, stochastic dynamics enable the network to address this by sampling from a posterior that combines these orthogonal bases.

\subsection{Stochastic retrieval as macro-scale Bayesian inference}

As a consequence of the \acrshort{fep}, the inference process described above - where each subparticle $\sigma_i$ updates its state based on local information (its bias $b_i$ and input $\sum_j J_{ij} \sigma_j$) - can be seen as a form of micro-scale inference, in which the prior - defined by the node's internal bias, gets updated by the evidence collected from the neighboring subparticles to form the posterior.
However, as the whole network itself is also a particular partition (specifically, a deep one), it must also perform Bayesian inference, at the macro-scale.
While the above argumentation provides a simple, self-contained proof, the nature of macro-scale inference can be elucidated by using the equivalence of the derived attractor network - in the special case of symmetric couplings - to Boltzmann machines (without hidden units). Namely, the ability of Boltzmann machines to perform macro-scale approximate Bayesian inference through Markov Chain Monte Carlo (MCMC) sampling has been well established in the literature \citep{ACKLEY_1985, Hinton_2002}.

Let us consider the network's learned weights $\mathbf{J}$ (and potentially its baseline biases $b^{\text{base}}$) as defining a \textbf{prior distribution} over the collective states $\boldsymbol{\sigma}$:
\begin{equation}
p(\boldsymbol{\sigma}) \propto \exp \Biggl\{ \sum_{i} b_i^{\text{base}} \sigma_i + \frac{1}{2} \sum_{ij} J_{ij}\sigma_i\sigma_j \Biggr\}
\end{equation}

Now, suppose the network receives external input (evidence) $\mathbf{s}$, which manifests as persistent modulations $\delta b_i$ to the biases, such that the total bias is $b_i = b_i^{\text{base}} + \delta b_i$. This evidence can be associated with a \textbf{likelihood function}:
\begin{equation}
p(\mathbf{s} | \boldsymbol{\sigma}) \propto \exp \left( \sum_i \delta b_i \sigma_i \right)
\end{equation}

According to Bayes' theorem, the \textbf{posterior distribution} over the network states given the evidence is $p(\boldsymbol{\sigma} | \mathbf{s}) \propto p(\mathbf{s} | \boldsymbol{\sigma}) p(\boldsymbol{\sigma})$:
\begin{equation}
\label{posterior-distribution}
p(\boldsymbol{\sigma} | \mathbf{s}) \propto \exp \Biggl\{ \sum_{i} b_i \sigma_i + \frac{1}{2} \sum_{ij} J_{ij}\sigma_i\sigma_j \Biggr\}
\end{equation}
As expected, this posterior distribution has the same functional form as the network's equilibrium distribution under the influence of the total biases $b_i$. Thus, the stochastic update rule derived from minimizing local \acrshort{vfe} (eq. (\ref{free-energy-rnn})) effectively performs Markov Chain Monte Carlo (MCMC) sampling - specifically Gibbs sampling - from the joint posterior distribution defined by the current \acrshort{vfe} landscape. In the presence of evidence $\mathbf{s}$, the network dynamics therefore \textit{sample} from the posterior distribution $p(\boldsymbol{\sigma} | \mathbf{s})$.
The significance of the stochasticity becomes apparent when considering the network's behavior over time. The time-averaged state $\langle \boldsymbol{\sigma} \rangle$ converges to the expected value under the posterior distribution:

\begin{equation}
\boxed{
\langle \boldsymbol{\sigma} \rangle = \lim_{T \to \infty} \frac{1}{T} \int_0^T \boldsymbol{\sigma}(t) dt \approx \mathbb{E}_{p(\boldsymbol{\sigma} | \mathbf{s})}[\boldsymbol{\sigma}]}
\end{equation}

Noise or stochasticity allows the system to explore the posterior landscape, escaping local minima inherited from the prior if they conflict with the evidence, and potentially mixing between multiple attractors that are compatible with the evidence. The resulting average activity $\langle \boldsymbol{\sigma} \rangle$ thus represents a Bayesian integration of the prior knowledge encoded in the weights and the current evidence encoded in the biases. This contrasts sharply with deterministic dynamics, which would merely settle into a single (potentially suboptimal) attractor within the posterior landscape.

Furthermore, if the learning process shapes the prior $p(\boldsymbol{\sigma})$ to have less redundant (e.g., more orthogonal) attractors, this parsimonious structure of the prior naturally contributes to a less redundant posterior distribution $p(\boldsymbol{\sigma} | \mathbf{s})$. When the prior belief structure is efficient and its modes are distinct, the posterior modes formed by integrating evidence $\mathbf{s}$ are also less likely to be ambiguous or highly overlapping. This leads to more robust and interpretable inference, as the network can more clearly distinguish between different explanations for the sensory data.

\begin{figure}[!htbp]
\centering
\includegraphics[width=1\linewidth]{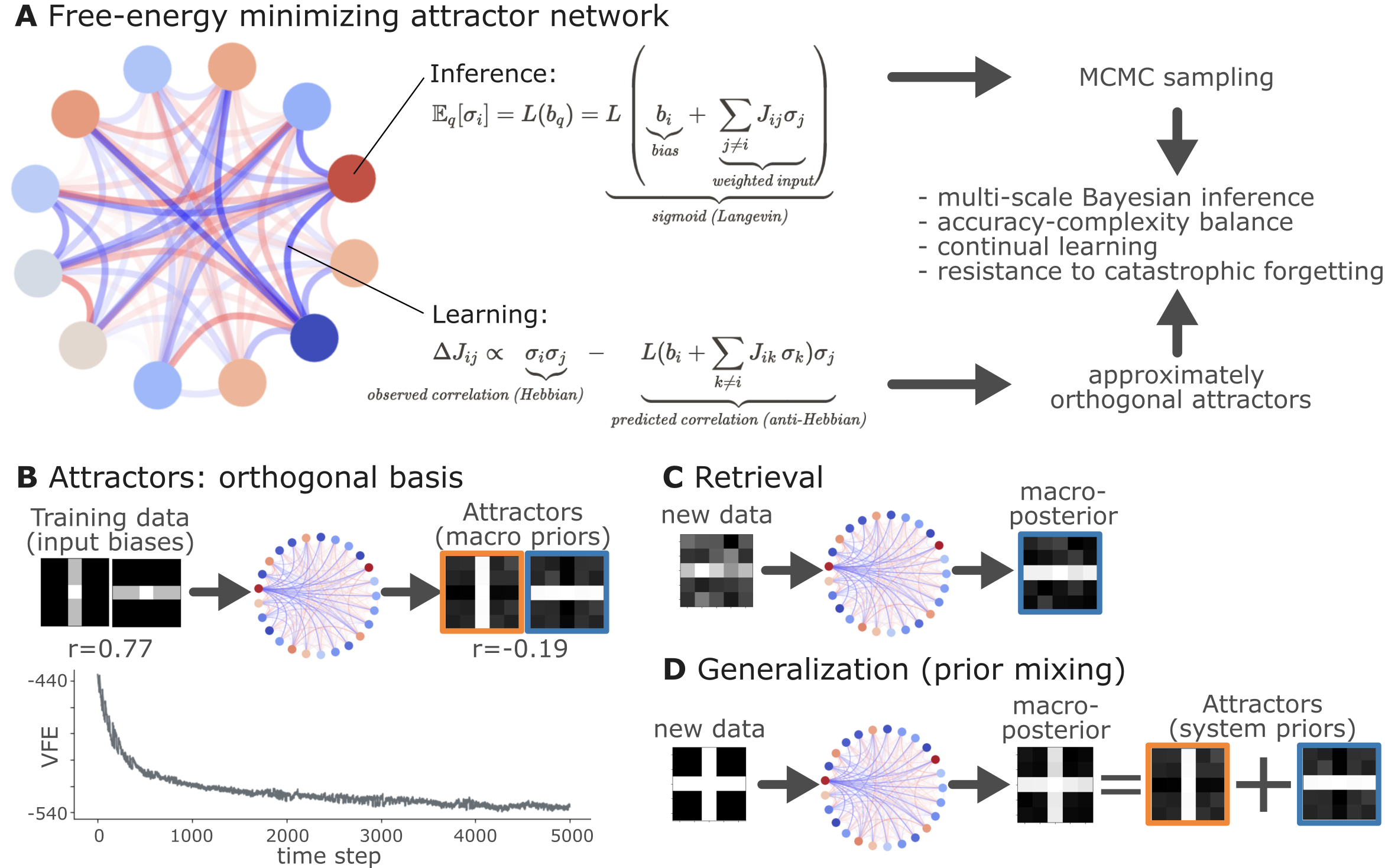}
\caption[]{\textbf{Free energy minimizing, adaptively self-organizing attractor network} \newline
\textbf{A} Schematic of the network illustrating inference and learning processes. Inference and learning are two faces of the same process: minimizing local variational free energy (\acrshort{vfe}), leading to dissipative dynamics and approximately orthogonal attractors.
\textbf{B} A demonstrative simulated example (\href{https://pni-lab.github.io/fep-attractor-network/simulation-demo}{Simulation 1}) of the network's attractors forming an orthogonal basis of the input data. Training can be performed by introducing the training data (top left) through the biases of the network. In this example, the input data consists of two correlated patterns (Pearson's r = 0.77). During repeated updates, micro-scale (local, node-level) \acrshort{vfe} minimization implements a simultaneous learning and inference process, which leads to approximate macro-scale (network-level) free energy minimization (bottom graph). The resulting network does not simply store the input data as attractors, but it stores approximately orthogonalized varieties of it (top right, Pearson's r = -0.19).
\textbf{C} When the trained network is introduced a noisy version of one of the training patterns (left), it is internally handled as the Likelihood function, and the network performs irreversible Markov-Chain Monte-Carlo (MCMC) sampling of the posterior distribution, given the priors defined by the network's attractors (top right), which can be understood as a retrieval process.
\textbf{D} Thanks to its orthogonal attractor representation, the network is able to generalize to new patterns - as long as they are sampled from the sub-space spanned by the attractors - by combining the quasi-orthogonal attractor states (bottom right) by multistable stochastic dynamics during the MCMC sampling.}
\label{fig-network}
\end{figure}

With this, the loop is closed. The stochastic attractor network emerging from the \acrshort{fep} framework (summarized on Figure~\ref{fig-network}A) naturally implements macro-scale Bayesian inference through its collective sampling dynamics, providing a robust mechanism for integrating prior beliefs with incoming sensory evidence --- as demonstrated concretely on the face recognition task (Figure~\ref{fig-scaling}F), where the network recovers face identity from heavily corrupted input by combining sensory evidence with attractor-based priors.
This reveals the potential for a deep recursive application of the Free Energy Principle: the emergent collective behavior of the entire network, formed by interacting subparticles each minimizing their local free energy, recapitulates the inferential dynamics of a single, macro-scale particle. This recursion could extend to arbitrary depths, giving rise to a hierarchy of nested particular partitions and multiple emergent levels of description, each level performing Bayesian active inference according to the same fundamental principles.

\subsection{Non-equilibrium steady state}
\label{self-solenoidization}

We now turn from the local update rules to the global steady-state behaviour of the network. The coupling matrix can always be decomposed as $\bm{J} = \bm{J}^{\dagger} + \bm{J}^ -$, where $\bm{J}^{\dagger} = \frac{1}{2}(\bm{J} + \bm{J}^\intercal)$ is symmetric and $\bm{J}^ - = \frac{1}{2}(\bm{J} - \bm{J}^\intercal)$ is antisymmetric. We consider these two components in turn.

\textbf{Symmetric couplings and the Boltzmann-like stationary distribution.} When the learning rule (eq. (\ref{learning-rule})) is driven by randomly ordered input patterns, it converges to approximately symmetric weights ($\bm{J} \approx \bm{J}^{\dagger}$). In this case the pairwise conditionals $p(\sigma_i | \sigma_j) \propto \exp\!\bigl((b_i + J_{ij}\sigma_j)\sigma_i\bigr)$ are symmetric in their coupling terms ($J_{ij} = J_{ji}$), and the conditional independence structure of the Markov blanket states $s_{ij}$, $a_{ij}$ \citep{Hipolito2021Markov} allows us to factorise the global joint as $p(\bm{\sigma}) \propto \prod_{i,j} p(\sigma_i, \sigma_j)$. Since $\sigma_i \sigma_j = \sigma_j \sigma_i$, one obtains:

\begin{equation}
\label{antisymmetric-coupling}
\sum_{i\neq j} J_{ij}\,\sigma_i\,\sigma_j = \sum_{i<j} \Bigl(J_{ij}+J_{ji}\Bigr)\,\sigma_i\,\sigma_j
\end{equation}

which, for symmetric $\bm{J}$, yields the Boltzmann form:

\begin{equation}
\label{hopfield-joint}
\boxed{
p(\bm{\sigma}) \propto \exp \Biggl\{ \underbrace{\underbrace{\sum_{i} b_i \sigma_i}_{\textit{bias term}} + \underbrace{ \sum_{ij} J^{\dagger}_{ij}\sigma_i\sigma_j}_{\textit{interaction term}} }_{\textit{-ve energy}} \Biggr\}
}
\end{equation}

with $\bm{J}^{\dagger} = \frac{1}{2} (\bm{J} + \bm{J}^\intercal)$ (and $J^{\dagger}_{ii} = 0$). This is the functional form of a stochastic continuous-state Hopfield network (a specific type of Boltzmann machine), with regions of high probability density constituting stochastic attractors.

\textbf{Asymmetric couplings and self-solenoidization.} When data arrives in fixed sequences, the learning rule additionally develops an antisymmetric component $\bm{J}^ -$. The simple factorisation argument above no longer applies, because directed couplings break the symmetry $J_{ij} \neq J_{ji}$. Nevertheless, we argue that the Boltzmann-like stationary distribution (eq. (\ref{hopfield-joint})) is approximately preserved, with solenoidal flows on top.

The key intuition is that self-orthogonalization provides the missing link. Because \acrshort{vfe}-driven learning pushes attractor representations toward approximate orthogonality, the antisymmetric coupling component $\bm{J}^ -$ --- which encodes directed transitions between attractors --- acts predominantly \textit{along} the iso-energy contours of the landscape defined by $\bm{J}^{\dagger}$. Any component of the antisymmetric flow that were to act \textit{across} iso-energy contours would imply shared information between attractors; but precisely this shared information is progressively absorbed into the symmetric part $\bm{J}^{\dagger}$ by the residual (anti-Hebbian) learning rule. In other words, the same mechanism that orthogonalizes attractors also ensures that the inter-attractor transition flows are approximately tangential to the energy landscape --- i.e., approximately solenoidal.

Under this condition, the network dynamics decompose into two complementary components: (i) \textbf{Dissipative (gradient) flow}, driven by $\bm{J}^{\dagger}$: descends the free energy landscape, pulling states toward attractor basins. Responsible for pattern storage and retrieval, and (ii) \textbf{Solenoidal (rotational) flow}, driven by $\bm{J}^ -$: generates probability currents along iso-energy contours --- analogous to a ball circling within a bowl rather than simply rolling to the bottom.

Thus the dynamics induced by $\bm{J}^{\dagger}$ and $\bm{J}^ -$ will approximately map to the generalized Helmholtz-Ao decomposition of the network dynamics (see also \href{https://pni-lab.github.io/fep-attractor-network/appendix/#appendix-5}{Appendix~5} and \citep{Friston_2023, Ao_2004, Xing_2010}). Because the solenoidal term transports probability mass along iso-energy surfaces without altering the potential, the stationary distribution retains the Boltzmann-like form of eq. (\ref{hopfield-joint}), while detailed balance is broken by persistent probability currents.

The practical consequences are twofold. First, directed sequence dynamics (heteroclinic chains) emerge naturally from the solenoidal component, enabling the network to traverse stored patterns in a learned temporal order --- as demonstrated in Simulation 3. Second, the non-reversible currents can transport probability mass between attractor basins without climbing energy barriers, potentially accelerating mixing relative to the reversible (symmetric-coupling) case \citep{Ao_2004, Xing_2010}.
A rigorous characterization of the relationship between attractor orthogonality and the divergence-free condition --- and of the resulting mixing-time improvements --- is an important direction for future work. As a first step, in the present work we perform simulations (\href{https://pni-lab.github.io/fep-attractor-network/simulation-digits-continuous-sequence}{Simulation 3}) to confirm that the key phenomenology --- stable attractors with directed inter-attractor transitions --- is robustly observed.

\section{In silico demonstrations}

We illustrate key features of the proposed framework with computer simulations. All simulations are based on python implementations of the network, available at \href{https://github.com/pni-lab/fep-attractor-network}{https://github.com/tspisak/fep-attractor-networks}. Simulations 1-4 are based on an implementation that favors clarity over efficiency - it implements both $\sigma$ and boundary states as separate classes, and is not optimized for performance. Simulations 5 and 6 are based on a more efficient vectorized implementation, to allow scaling up to more complex data.
In all simulations, we train an attractor network with the derived rules in a continuous-learning fashion (i.e simultaneously performing inference and learning). To be able to control the precision during inference and the speed of learning, we introduce two coefficients for eq.-s (\ref{inference-rule}) and (\ref{learning-rule}), the inverse temperature parameter $iT$ and a learning-rate $\alpha$.

\subsection{Simulation 1: demonstration of orthogonal basis formation, macro-scale free energy minimization and Bayesian inference}

In \href{https://pni-lab.github.io/fep-attractor-network/simulation-demo}{Simulation 1}, we construct a network with 25 subparticles (representing 5x5 images) and train it with 2 different, but correlated images (Pearson's r = 0.77, see Figure~\ref{fig-network}B), with a precision of 0.1 and a learning rate of 0.01 (see next simulation for parameter-dependence). The training phase consisted of 500 epochs, each showing a randomly selected pattern from the training set through 10 time steps of simultaneous inference and learning. As shown on Figure~\ref{fig-network}B, local, micro-scale \acrshort{vfe} minimization performed by the simultaneous inference and learning process leads to a macro-scale free energy minimization. Next, we obtained the attractor states corresponding to the input patterns by means of deterministic inference (updating with the expected value, instead of sampling from the $\mathcal{CB}$ distribution, akin to a vanilla Hopfield network). As predicted by theory, the attractor states were not simple replicas of the input patterns, but approximately orthogonalized versions of them, displaying a correlation coefficient of r=-0.19.
Next, we demonstrated that the network (with stochastic inference) is not only able to retrieve the input patterns from noisy variations of them (fig. Figure~\ref{fig-network}C), but also generalizes well to reconstruct a third pattern, by combining its quasi-orthogonal attractor states (fig. Figure~\ref{fig-network}D). Note that this simulation only aimed to demonstrate some of the key features of the proposed architecture, and a comprehensive evaluation of the network's performance, and its dependency on the parameters is presented in the next simulation.

\subsection{Simulation 2: systematic evaluation of learning regimes}

In \href{https://pni-lab.github.io/fep-attractor-network/simulation-digits}{Simulation 2}, we trained the network on 10 images of handwritten digits (a single example of each of the 10 digits from 0 to 9, 8x8 pixels each, as distributed with scikit-learn, see Figure~\ref{fig-digits}C, upper row). The remaining 1787 images were unseen in the training phase and only used as a test set, in a one-shot learning fashion. The network was trained with a fixed learning rate of 0.01, through 5000 epochs, each consisting of 10 time steps with the same, randomly selected pattern from the training set of 10 images, while performing simultaneous inference and learning. We evaluated the effect of the inverse temperature parameter $iT$ (i.e. precision) and the strength of evidence during training, i.e. the magnitude of the bias changes $\delta b_i$.
The precision parameter $iT$ was varied with 19 values between 0.01 and 1, and the strength of evidence during training varied by changing the magnitude of the biases from 1 to 20, with increments of 1. The training patterns were first preprocessed by squaring the pixel values (to enhance contrast) and normalizing each image to have zero mean and unit variance. We performed a total of 380 runs, varying these parameters in a grid-search fashion. All cases were evaluated in terms of (i) stochastic (Bayesian) pattern retrieval from noisy variations of the training images; and (ii) one-shot generalization to reconstruct unseen handwritten digit examples. In both types of evaluation, the network was presented a noisy variant of a randomly selected (training or test) image through its biases. The noisy patterns were generated by adding Gaussian noise with a standard deviation of 1 to the pixel values of the training images (see ``Examples'' in Figure~\ref{fig-digits}B C and D). The network's response was obtained by averaging 100 time steps of stochastic inference. The performance was quantified as the improvement in the proportion of variance in the original target pattern (without noise) explained by the network's response, compared to that explained by the noisy input pattern. Both for retrieval and generalization, this approach was repeated 100 times, with a randomly sampled image from the training (10 images) and test set (1787 images), respectively. The median improvement across these 100 repetition was used as the primary performance metric. The retrieval and 1-shot generalization performance of models trained with different $iT$ and $\alpha$ parameters is shown on Figure~\ref{fig-digits}A, top row). We found that, while retrieval of a noisy training pattern was best with precision values between 0.1 and 0.5, generalization to new data preferred lower precision during learning ($iT$\textless 0.1), i.e. more stochastic dynamics.
Furthermore, in all simulation cases, we seeded the networks with the original test patterns and obtained the corresponding attractor states, by means of deterministic inference. We then computed the pairwise correlation and dot product between the attractor states. The dot product was converted to degrees. Orthogonality was finally quantified by the mean correlation among attractors and the mean squared deviation from orthogonality (in degrees). To establish a reference value, the same procedure was also repeated for the original patterns (after preprocessing), which displayed a mean correlation a 29.94 degree mean squared deviation from orthogonality. Attractor orthogonality and the number of attractors for each simulation case is shown on Figure~\ref{fig-digits}A, bottom row.
We found that depending on the temperature of the network during the learning phase, the network can be in characteristic regimes of high, low and balanced complexity (Figure~\ref{fig-digits}B). With low temperature (high precision), high model complexity is allowed (``accuracy pumping'') and attractors will tend to exactly match the training data Figure~\ref{fig-digits}C. On the contrary, high temperatures (low precision) result in a single fixed point attractor and reduced recognition performance Figure~\ref{fig-digits}E. However, such networks were found to be able to generalize to new data, suggesting the existence of ``soft attractors'' (e.g. saddle-like structures) that are not local minima on the free energy landscape, yet affect the steady-state posterior distribution in a non-negligible way (especially with longer mixing-times). A balanced regime Figure~\ref{fig-digits}D can be found with intermediate training precision, where both recognition and 1-shot generalization performance are high (similarly to the ``standard regime'' of \textit{prospective configuration} \citep{https://doi.org/10.48550/arxiv.2207.12316}). This is exactly the regime that promotes attractor orthogonalization, crucial for efficient representation and generalization. The complexity restrictions on these models cause them to re-use the same attractors to represent different patterns (see e.g. the single attractor belonging to the digits 5 and 7 in the example on panel D), which eventually leads to less attractors, with each having more explanatory power, and being approximately orthogonal to each other. Panels C-E on Figure~\ref{fig-digits} provide examples of network behavior on a handwritten digit task across different regimes, including (i) training data (same in all cases); (ii) fixed-point attractors (obtained with deterministic update); (iii) attractor-orthogonality (polar histogram of the pairwise angles between attractors); (iv) retrieval and 1-shot generalization performance ($R^2$ between the noisy input pattern and the network output after 100 time steps, for 100 randomly sampled patterns) and (v) illustrative example cases from the recognition and 1-shot generalization tests (noisy input, network output and true pattern).

\begin{figure}[!htbp]
\centering
\includegraphics[width=1\linewidth]{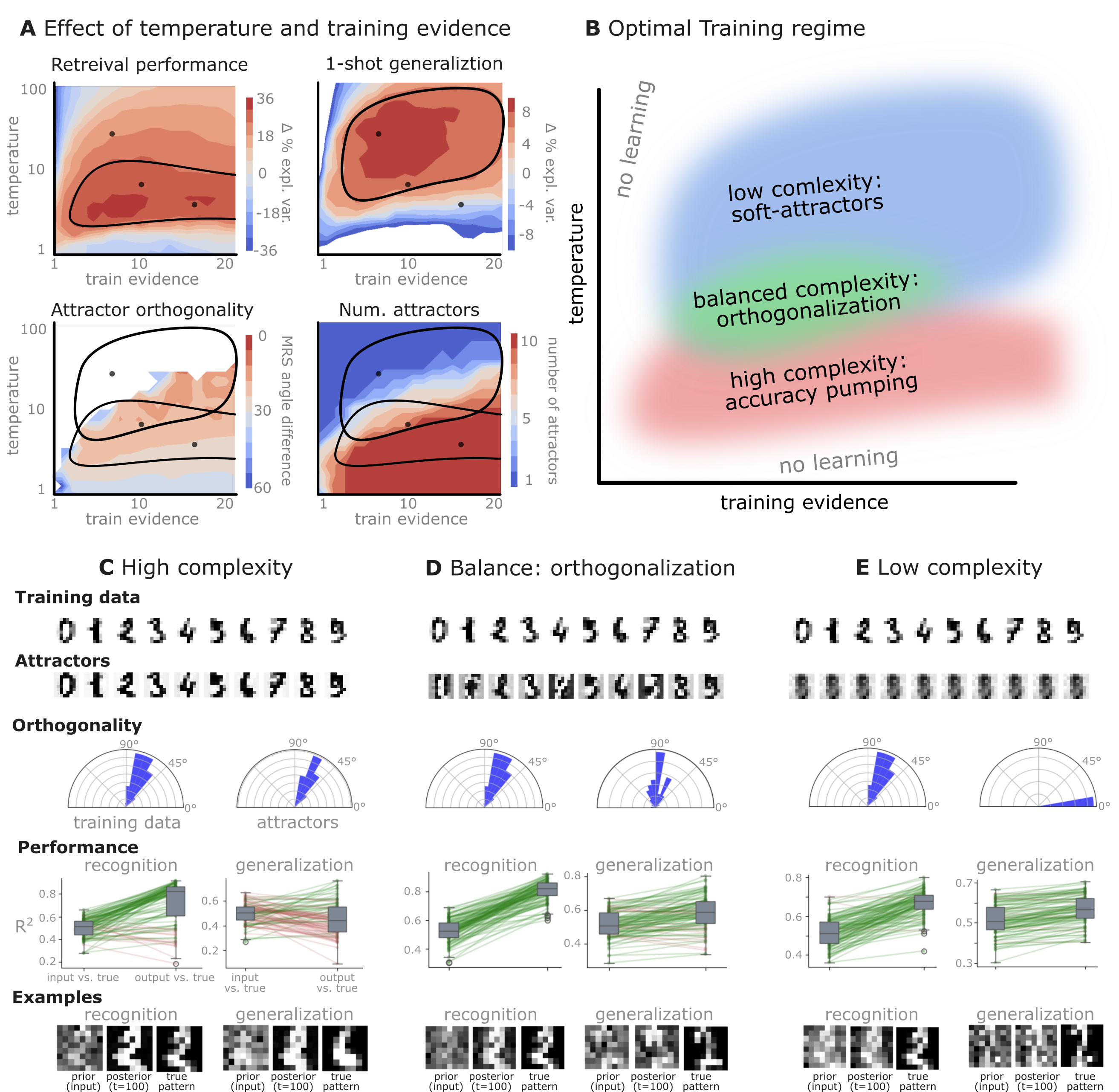}
\caption[]{\textbf{Adaptive self-organization and generalization in a free-energy minimizing attractor network.} \newline
Simulation results from training the network on a single, handwritten example for each of the 10 digits (0-9), with variations in training precision and evidence strength to explore different learning regimes (\href{https://pni-lab.github.io/fep-attractor-network/simulation-digits}{Simulation 2}).
\textbf{A}: Performance landscapes as a function of inference temperature (inverse precision) and training evidence strength (bias magnitude). Retrieval performance (reconstructing noisy variants of the 10 training patterns, top left), one-shot generalization (reconstructing a noisy variants of unseen handwritten digits, top right), attractor orthogonality (mean squared angular difference from 90° indicating higher orthogonality for lower values, bottom left), and the number of attractors (when initialized with the 10 training patterns, bottom right) are shown. Optimal regions (contoured) highlight parameter settings that yield good generalization and highly orthogonal attractors. Contours in the top left and top right highlight the most efficient parameter settings for retrieval and generalization, respectively. Both contours are overlaid on the two bottom plots.
\textbf{B}: Conceptual illustration of training regimes. With low temperature (high precision) high model complexity is allowed (``accuracy pumping'') and attractors will tend to exactly match the training data. On the contrary, high temperatures (low precision) result in a single fixed point attractor and reduced recognition performance. However, such networks will be able to generalize to new data, suggesting the existence of ``soft attractors'' (e.g. saddle-like structures) that are not local minima on the free energy landscape, yet affect the steady-state posterior distribution in a non-negligible way (especially with longer mixing-times).}
\label{fig-digits}
\end{figure}

\begin{figure}
  \ContinuedFloat
  \caption{(continued) A balanced regime can be found with intermediate precision during training, where both recognition and generalization performance are high. This is exactly the regime that promotes attractor orthogonalization, crucial for efficient representation and generalization. The complexity restrictions on these models cause them to re-use the same attractors to represent different patterns (see e.g. the single attractor belonging to the digits 5 and 7 in the example on panel D), which eventually leads to approximate orthogonality. Panels C-E provide examples of network behavior on a handwritten digit task across different regimes, including (i) training data (same in all cases); (ii) fixed-point attractors (obtained with deterministic update); (iii) attractor-orthogonality (polar histogram of the pairwise angles between attractors); (iv) retrieval and 1-shot generalization performance ($R^2$ between the noisy input pattern and the network output after 100 time steps, for 100 randomly sampled patterns) and (v) illustrative example cases from the recognition and 1-shot generalization tests (noisy input, network output and true pattern).
\textbf{C}: High complexity: Attractors are sharp and similar to training data; good recognition, limited generalization.
\textbf{D}: Balanced complexity (orthogonalization): Attractors are distinct and quasi-orthogonal, enabling strong recognition and generalization from noisy inputs. The balanced regime clearly demonstrates the network's ability to form an orthogonal basis, facilitating effective generalization as predicted by the free-energy minimization framework.
\textbf{E}: Low complexity: There is only a single fixed-point attractor. Recognition performance is lower, but generalization remains considerable.}
\end{figure}

\subsection{Simulation 3: demonstration of sequence learning capabilities}

\begin{figure}[!htbp]
\centering
\includegraphics[width=1\linewidth]{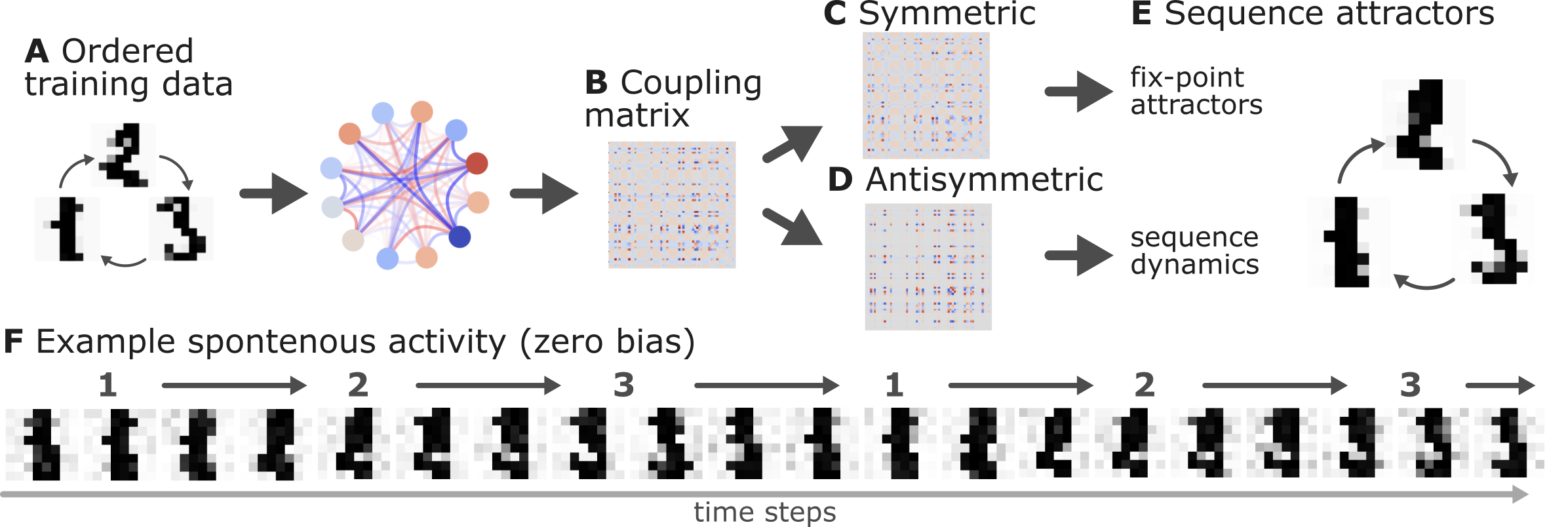}
\caption[]{\textbf{Sequential Dynamics in Free-Energy Minimizing Attractor Networks.} \newline

Simulation results (\href{https://pni-lab.github.io/fep-attractor-network/simulation-digits-continuous-sequence}{Simulation 3}) demonstrate the framework's ability to learn temporal sequences. Ordered training data leads to asymmetric coupling matrices, where the symmetric component establishes fixed-point attractors for individual patterns, and the antisymmetric component encodes the transitional dynamics, enabling spontaneous sequence recall.
\textbf{A}: Ordered training data (digits 1, 2, 3) presented sequentially to the network.
\textbf{B}: The emergent coupling matrix, displaying asymmetry as a consequence of sequential training.
\textbf{C}: The symmetric component of the coupling matrix. This part is responsible for creating stable, fixed-point attractors for each pattern in the sequence.
\textbf{D}: The antisymmetric component of the coupling matrix. This part drives the directional transitions between the attractors, encoding the learned order.
\textbf{E}: The sequence attractors themselves -- fixed-point attractors (digits 1, 2, 3) derived from the symmetric coupling component.
\textbf{F}: Example of spontaneous network activity with zero external bias, showcasing the autonomous recall of the learned sequence $(1 \rightarrow2 \rightarrow3 \rightarrow...)$ driven by the interplay of symmetric and antisymmetric couplings.}
\label{fig-sequence-attractors}
\end{figure}

In \href{https://pni-lab.github.io/fep-attractor-network/simulation-digits-continuous-sequence}{Simulation 3}, we demonstrate the sequence learning capabilities of the proposed architecture. We trained the network on a sequence of 3 handwritten digits (1,2,3), with a fixed order of presentation ($1 \rightarrow2 \rightarrow3 \rightarrow1 \rightarrow2 \rightarrow3 \rightarrow...$), for 2000 epochs, each epoch consisting of a single step (Figure~\ref{fig-sequence-attractors}A). This rapid presentation of the input sequence forced the network to model the current attractor from the network's response to the previous pattern, i.e. to establish sequence attractors. The inverse temperature was set to 1 and the learning rate to 0.001 (in a supplementary analysis, we saw a considerable robustness of our results to the choice of these parameters). As shown on Figure~\ref{fig-sequence-attractors}B, this training approach led to an asymmetric coupling matrix (it was very close to symmetric in all previous simulations). Based on eq.-s (\ref{antisymmetric-coupling}) and (\ref{hopfield-joint}), we decomposed the coupling matrix into a symmetric and antisymmetric part (Figure~\ref{fig-sequence-attractors}C and D). Retrieving the fixed-point attractors for the symmetric component of the coupling matrix, we obtained three attractors, corresponding to the three digits (Figure~\ref{fig-sequence-attractors}C and E). The antisymmetric component of the coupling matrix, on the other hand, was encoding the sequence dynamics. Indeed, letting the network freely run (with zero bias) resulted in a spontaneously emerging sequence of variations of the digits $1 \rightarrow2 \rightarrow3 \rightarrow1 \rightarrow2 \rightarrow3 \rightarrow1 \rightarrow...$, reflecting the original training order (Figure~\ref{fig-sequence-attractors}F). This illustrates that the proposed framework is capable of producing and handling asymmetric couplings, and thereby learn sequences.

\subsection{Simulation 4: demonstration of resistance to catastrophic forgetting}

\begin{figure}[!htbp]
\centering
\includegraphics[width=0.66\linewidth]{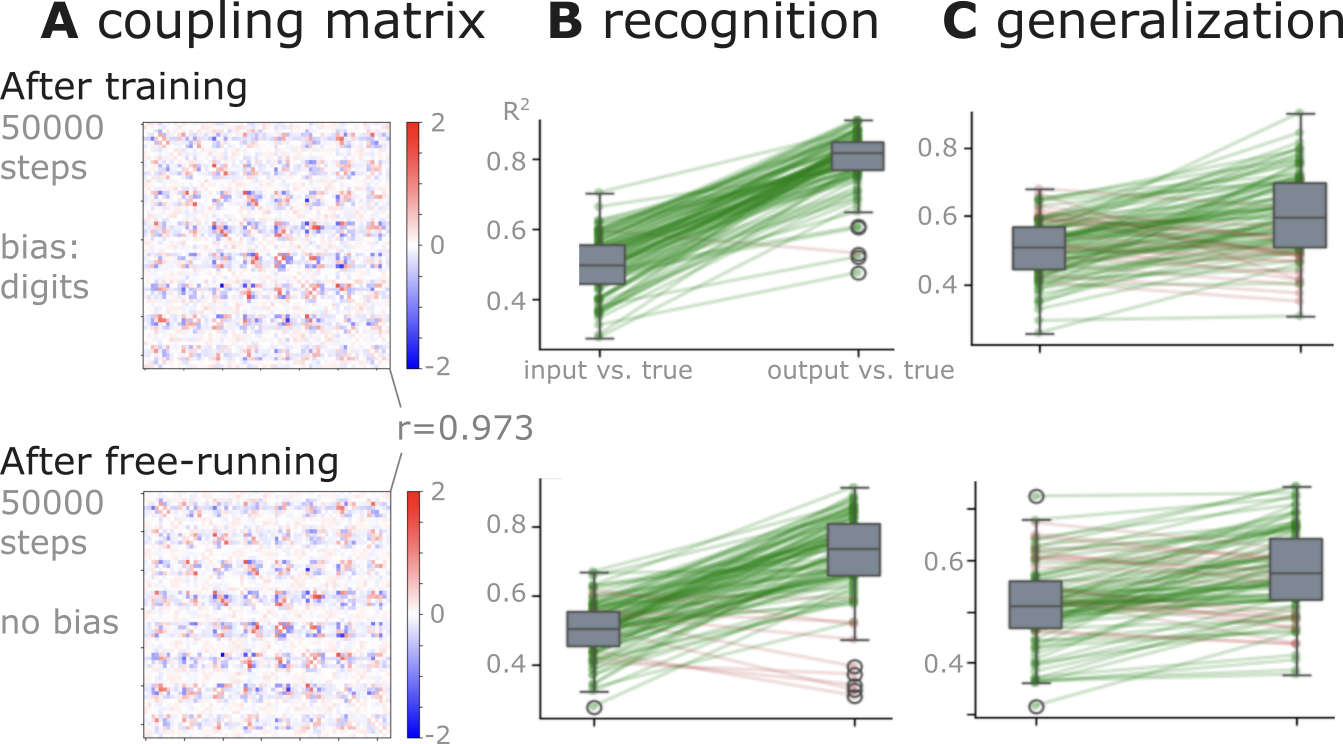}
\caption[]{\textbf{Demonstration of resistance to catastrophic forgetting via spontaneous activity.}  \newline

Simulation results (\href{https://pni-lab.github.io/fep-attractor-network/simulation-digits-catastrophic-forgetting}{Simulation 4}) illustrate the network's ability to mitigate catastrophic forgetting. When allowing the network to ``free-run'' (e.g. with zero external bias) while performing continuous weight adjustment, spontaneous activity reinforces existing attractors, largely preserving learned knowledge even in the absence of the repeated presentation of previous training patterns.
\textbf{A}: Coupling matrices. The top panel displays the coupling matrix immediately after training on digit patterns (50000 steps with bias corresponding to digits, the same simulation case as on Figure~\ref{fig-digits}D). The bottom panel shows the coupling matrix after an additional 50000 steps of free-running (zero bias, but active weight adjustment), indicating that the learned structure is largely maintained.
\textbf{B}: Recognition performance. This panel compares the $R^2$ values for reconstructing noisy versions of the trained digit patterns. It shows the similarity between the noisy input and the true pattern (left boxplots in each sub-panel) versus the similarity between the network's output and the true pattern (right boxplots in each sub-panel). Performance is robustly maintained after the free-running phase (bottom) compared to immediately after training (top).
\textbf{C}: One-shot generalization performance. This panel shows the $R^2$ values for reconstructing noisy versions of \textit{unseen} handwritten digit patterns (after seeing only a single example per digit). Similar to recognition, the network's ability to generalize to novel inputs is well-preserved after the free-running phase (bottom) compared to immediately after training (top).}
\label{fig-catastrophic-forgetting}
\end{figure}

In \href{https://pni-lab.github.io/fep-attractor-network/simulation-digits-catastrophic-forgetting}{Simulation 4}, we took a network trained in Simulation 2 with inverse temperature 0.17 and evidence level 11 (the same network that is illustrated on Figure~\ref{fig-digits}D) and let it run for 50000 epochs (the same number of epochs as the training phase), with unchanged learning rate, but zero bias. We expected that, as the network spontaneously traverses around its attractors, it reinforces them and prevents them from being fully ``forgotten''. Indeed, as shown on Figure~\ref{fig-catastrophic-forgetting}, the network's coupling matrix (panel A), retrieval performance (panel B) and one-shot generalization performance (panel C) were all very similar to the original network's performance. However, the network's attractor states were not exactly the same as the original ones, indicating that some of the original attractors become ``soft attractors'' (or ``ghost attractors''), which do not represent explicit local minima on the free energy landscape anymore, but their influence on the network's dynamics is still significant (see \href{https://pni-lab.github.io/fep-attractor-network/simulation-digits-catastrophic-forgetting}{Simulation 4}).

\subsection{Simulation 5: scalability profile and memory capacity}

In addition to the reference Python implementation used for Simulations 1--4, we provide a vectorized JAX implementation that applies the same inference and learning rules (eq. (\ref{inference-rule}), (\ref{learning-rule})) in a parallelized, full-network update schedule (\href{https://pni-lab.github.io/fep-attractor-network/simulation-scaling-jax}{Simulation 5}). This parallel schedule is computationally more efficient but not strictly equivalent to the sequential node-by-node updates of the reference implementation; we validate that both implementations produce qualitatively similar coupling matrices and retention behavior on shared test cases. The JAX implementation is used exclusively for runtime profiling (this simulation) and for the larger-scale face recognition experiment (Simulation 6).

We profile the computational scaling of the JAX implementation across network sizes ranging from $N = 64$ to $N = 50000$ nodes, i.e. up to 2.5 billion total parameters (\href{https://pni-lab.github.io/fep-attractor-network/simulation-scaling-jax}{Simulation 5}; see also \href{https://pni-lab.github.io/fep-attractor-network/appendix/#appendix-9}{Appendix~9} for detailed benchmarks). Training and inference runtimes scale as expected from the analytical $O(N^2)$ per-step complexity (Figure~\ref{fig-scaling}A). Throughput remains high for moderate network sizes and drops off as expected for very large $N$ (Figure~\ref{fig-scaling}B). We further probe memory capacity by varying the number of stored patterns $K$ at fixed network size, measuring both retention (deterministic attractor recovery, Figure~\ref{fig-scaling}C) and noisy reconstruction quality (Figure~\ref{fig-scaling}D). Consistent with the analytical prediction that emergent orthogonalization approaches the projector-network capacity limit, the network maintains high-fidelity attractors and effective Bayesian retrieval for pattern counts substantially exceeding the classical Hopfield bound of $\sim 0.14N$.

\subsection{Simulation 6: face recognition with the Olivetti faces dataset}

\begin{figure}[!htbp]
\centering
\includegraphics[width=1\linewidth]{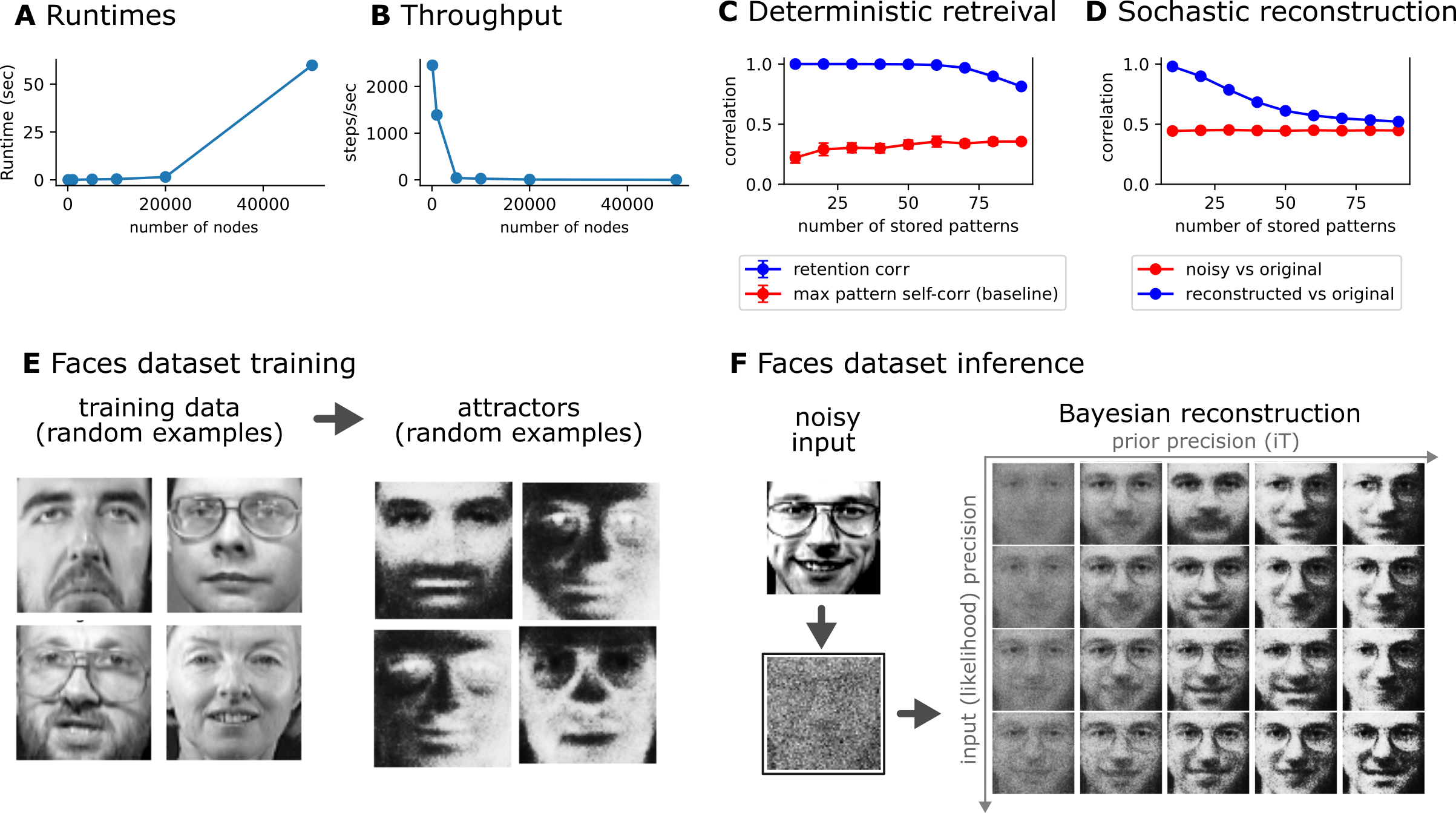}
\caption[]{\textbf{Scaling, memory capacity, and face recognition.} \newline

\textbf{A--B}: Runtime (A) and throughput (B) of the JAX implementation as a function of network size $N$, confirming the expected $O(N^2)$ per-step scaling (\href{https://pni-lab.github.io/fep-attractor-network/simulation-scaling-jax}{Simulation 5}).
\textbf{C}: Deterministic retrieval quality (retention correlation, blue) as a function of the number of stored patterns $K$, compared with the maximum inter-pattern self-correlation baseline (red). The network maintains high-fidelity attractors well beyond the classical Hopfield capacity bound.
\textbf{D}: Stochastic (Bayesian) reconstruction: correlation of the reconstructed output with the original pattern (blue) versus correlation of the noisy input with the original (red), as a function of $K$. The network consistently improves upon the noisy input, demonstrating effective Bayesian retrieval.
\textbf{E}: Training the network on the Olivetti faces dataset ($64 \times 64$, $N = 4096$, 400 patterns; \href{https://pni-lab.github.io/fep-attractor-network/simulation-faces-jax}{Simulation 6}). Random examples of training faces (left) and the corresponding learned attractors (right), confirming attractor formation and orthogonalization in this naturalistic domain.
\textbf{F}: Bayesian face reconstruction from noisy input. Columns show reconstructions at increasing likelihood precision (left to right, top row) and increasing prior precision (top to bottom). Even when the input is degraded to a level approaching the limit of human face recognition, the network recovers the original identity by combining sensory evidence with its learned attractor-based priors.}
\label{fig-scaling}
\end{figure}

To move beyond the small-scale handwritten digit experiments, we test the framework on the Olivetti faces dataset --- a benchmark originally created at AT\&T Laboratories Cambridge, consisting of 400 grayscale face images (40 subjects, 10 images each) at full resolution $64 \times 64$ pixels ($N = 4096$ nodes, i.e. more than 16 million parameters to optimize), using the full set of 400 patterns (\href{https://pni-lab.github.io/fep-attractor-network/simulation-faces-jax}{Simulation 6}). Training confirms that the key phenomena observed on handwritten digits --- emergent orthogonalization and attractor formation --- transfer to this substantially higher-dimensional and more naturalistic stimulus domain (Figure~\ref{fig-scaling}E). We further evaluate stochastic reconstruction from heavily corrupted input. As shown on Figure~\ref{fig-scaling}F, the network performs Bayesian inference by integrating the noisy sensory evidence (likelihood) with its learned prior beliefs (attractors), with the balance controlled by the precision (inverse temperature) parameter. Notably, the input noise level used here is severe enough that the corrupted faces approach the limit of human recognizability, yet the network reliably recovers the identity of the original face, illustrating the power of Bayesian reconstruction via attractor-based priors.

\section{Discussion}

The crucial role of attractor dynamics in elucidating brain function mandates a foundational question: what types of attractor networks arise naturally from the first principles of self-organization, as articulated by the Free Energy Principle \citep{Friston_2023, Friston_2010}? Here we aimed to address this question. We demonstrated mathematically, by recourse to a prototypical parametrization, that the ensuing networks generally manifest as non-equilibrium steady-state (\acrshort{ness}) systems \citep{Xing_2010, Ao_2004} that have a stationary state probability distribution governed by the symmetric component of their synaptic efficacies, conforming to a Boltzmann-like form \citep{Amit_1989, Hochreiter_1997}. This renders the resulting self-organizing attractor networks a generalization of canonical, single-layer Boltzmann machines or stochastic Hopfield networks \citep{Hinton_2002, Hopfield_1982}, but distinguished by their capacity for asymmetric coupling and continuous-valued neuronal states (see \href{https://pni-lab.github.io/fep-attractor-network/appendix/#appendix-8}{Appendix~8} for a detailed comparison). The main assumptions underpinning this derivation - namely, the existence of a (deep) particular partition and the ensuing imperative to minimize variational free energy - are not merely parsimonious, but arguably fundamental for any random dynamical system that maintains its integrity despite a changing environment \citep{Friston_2023}.

Our formulation reveals that the minimization of variational free energy at the micro-scale, i.e., by individual network nodes (``subparticles''), gives rise to a dual dynamic within the active inference framework \citep{Friston_2009}. Firstly, it prescribes Bayesian update dynamics for the individual network nodes, homologous to the stochastic relaxation observed in Boltzmann architectures (e.g. stochastic Hopfield networks), with high neuroscientific relevance, especially for coarse-grained brain networks, where activity across brain regions have been reported to ``flow'' following a similar rule \citep{Cole_2016, Sanchez_Romero_2023, https://doi.org/10.48550/arxiv.2402.02191}. Secondly, it engenders a distinctive coupling plasticity - a local, incremental learning rule - that continuously adjusts coupling weights to preserve low free energy in anticipation of future sensory encounters, effectively implementing action selection in the active inference sense \citep{Friston_2016}.

The learning rule itself displays a high neurobiological plausibility, as it resembles both generalized Hebbian-anti-Hebbian learning \citep{F_ldi_k_1990, Sanger_1989} and predictive coding \citep{Rao_1999, https://doi.org/10.48550/arxiv.2202.09467, Millidge_2024}.
The neurobiological mechanisms of synaptic plasticity are best understood and experimentally most thoroughly studied at the level of single neurons, although efforts exist to scale up these findings to the level of populations of neurons (e.g., \citet{Robinson_2011, Huang_2023}). We believe that our framework holds promise for a better understanding of plasticity mechanisms independent of scale, as it mathematically survives arbitrary coarse-graining under the deep particular partition formalism.
At the single-neuron limit, the rule reduces to a discrete-time binary Hebbian/anti-Hebbian update (formally recovered when precision $\rightarrow \infty$), closely resembling spike-timing--dependent plasticity (STDP)  \citep{Roberts2010Anti, Vignoud_2024}, where correlated activity produces long-term potentiation (LTP {\textasciitilde} Hebbian term) and predicted activity leads to long-term depression (LTD {\textasciitilde} anti-Hebbian term; see e.g. \citet{Caporale_2008}). Thereby, our framework connects STDP to predictive coding \citep{Saponati_2023}, in a sense that presynaptic activity that is reliably predicted by postsynaptic firing is eventually depressed.\newline
Moreover, the subtractive predictive error term, together with the bounded nature of the continuous Bernoulli distribution, prevent runaway potentiation, functioning analogously to homeostatic and metaplasticity mechanisms \citep{Turrigiano_2011, Zenke_2017}.
Also similarly to biologically inspired models of plasticity, the \acrshort{fep}-based learning rule efficiently implements sequence learning, by contrasting the current correlations of inputs with that predicted by the network's generative model in the previous time step.
In sum, rather than fitting neuron-level data post hoc, our framework predicts that Hebbian/anti-Hebbian-style updates should appear at all descriptive scales, with differences only in implementation, not mathematical structure. The biological literature at synaptic, circuit, and network levels is largely consistent with this multiscale interpretation.

Furthermore, by capitalizing on its natural tendency to spontaneously (stochastically) revisit its attractors, the network is able to mitigate \textit{catastrophic forgetting} \citep{https://doi.org/10.48550/arxiv.2312.10549} - the tendency of current deep learning architectures to lose old representations when learning new ones. This phenomenon may serve as a model for the spontaneous fluctuations observed during resting state brain activity (also related to ``day-dreaming'') and shows promise for leveraging similar mechanisms in artificial systems.

Importantly, we show that in our adaptive self-organizing network architecture, micro-scale free energy minimization manifests in (approximate) macro-scale free energy minimization - as expected from the fact that the macro-scale network itself is also a particular partition. This entails that the network performs Bayesian inference not only locally in its nodes, but also on the macro-scale - a result well known from the literature on Boltzmann machines and also spiking neural networks \citep{ACKLEY_1985, Hinton_2002, Buesing_2011}. Our work extends these previous results with a holistic view: the free energy landscape, sculpted by the coupling efficacies and manifested as the repertoire of (soft-) attractors, encodes the system's \textit{prior beliefs}. Sensory inputs or internal computations, presented as perturbations to the internal biases of network nodes, constitute the \textit{likelihood}, while the network's inherent stochastic dynamics explore the \textit{posterior} landscape via a process akin to irreversible Markov Chain Monte Carlo sampling \citep{Gelman_1992}. The stochastic nature of these dynamics is not a mere epiphenomenon; it empowers the network to engage in active inference by dynamically traversing trajectories in its state space that blend the contributions of disparate attractor basins \citep{Friston_2009}. Consequently, for inputs that, while novel, reside within the subspace spanned by the learned attractors, the network generalizes by engaging in oscillatory activity - a characteristic signature of neural computations in the brain. While the notion of oscillations being generated by multistable stochastic dynamics has been previously proposed by e.g. \citep{Liu_2022}, the quasi-orthogonal basis spanned by the attractors of the system could render such a mechanism especially effective.

The \acrshort{fep} formalisation via deep particular partitions intrinsically accommodates multiple, hierarchically nested levels of description. One may arbitrarily coarse-grain the system by combining subparticles, ultimately arriving at a simple particular partition. Drawing upon concepts such as the center manifold theorem \citep{Wagner_1989}, it is posited that rapid, fine-grained dynamics at lower descriptive levels converge to lower-dimensional manifolds, upon which the system evolves via slower processes at coarser scales. This inherent separation of temporal scales offers a compelling paradigm for understanding large-scale brain dynamics, where fast neuronal activity underwrites slower cognitive processes through hierarchical active inference \citep{Man_2018}.
Indeed, empirical investigations have provided manifold evidence for attractor dynamics in large-scale brain activity \citep{Rolls_2009, Kelso_2012, Haken_1978, Breakspear_2017, Deco_2012, Kelso_2012, Gosti_2024, Chen_2025}.

A pivotal outcome of the \acrshort{fep}-driven simultaneous learning and inference process is the propensity of the network to establish \textit{approximately orthogonal attractor states}. We argue that this remarkable property is not simply a side effect; it is an unavoidable result of minimizing the variational free energy of conservative particles \citep{Friston_2023}; which entails minimizing model complexity and maximizing accuracy simultaneously or - equivalently - minimize redundancy via a maximization of mutual information. This key characteristic makes free energy minimizing attractor networks naturally approximate one of the most efficient attractor network architectures, the \textit{projector neural networks} articulated by Kanter and Sompolinsky \citep{Kanter_1987, Personnaz_1985}. We propose the approximate orthogonality of attractor states as a signature of free energy minimizing attractor networks, potentially detectable in natural systems (e.g. neural data).
Moreover, self-orthogonalization plays a structural role beyond efficient representation: by ensuring that attractors are approximately independent, it guarantees that the antisymmetric coupling component --- which encodes sequence dynamics and solenoidal flows --- acts approximately tangentially to the energy landscape defined by the symmetric part. This clean separation may point to a deeper formal correspondence between the two components of the coupling matrix and the two complementary objectives of the \acrshort{fep}: variational free energy minimization (perception, encoded in the symmetric component) and expected free energy minimization (action and planning, encoded in the antisymmetric component) --- an important direction for future theoretical work.
There is initial evidence from large-scale brain network data pointing into this direction. Using an attractor network model very similar to the herein presented - a recent paper has demonstrated that standard resting-state networks (RSNs) are manifestations of brain attractors that can be reconstructed from fMRI brain connectivity data \citep{Englert_2025}. Most importantly, these empirically reconstructed large-scale brain attractors were found to be largely orthogonal - the key feature of the self-organizing attractor networks described here. Future research needs to carefully check if these large-scale brain attractors, and the computing abilities that come with them, are truly a direct signature of an underlying \acrshort{fep}-based attractor network and how they are connected with related frameworks of ``dreaming neural networks'' \citep{Hopfield_1983, Plakhov, Dotsenko_1991, Fachechi_2019}, implementing similar mechanisms.

Besides neuroscientific relevance, our work has also manifold implications for artificial intelligence research. Relative to classical single-layer Hopfield/Boltzmann formulations, the present framework preserves attractor-based inference while extending the model class along several dimensions that emerge from the \acrshort{fep} derivation rather than being imposed by design. These extensions --- and their analytical consequences --- are elaborated in \href{https://pni-lab.github.io/fep-attractor-network/appendix/#appendix-8}{Appendix~8}. Two points deserve emphasis here. First, the learning rule (eq. (\ref{learning-rule})) eliminates the costly free-running phase of contrastive divergence, reducing per-step learning complexity from $O(N^2 k)$ to $O(N^2)$ (\href{https://pni-lab.github.io/fep-attractor-network/appendix/#appendix-8}{Appendix~8}). Second, in the presence of asymmetric couplings the solenoidal component of the dynamics induces non-reversible probability currents that accelerate mixing relative to symmetric Boltzmann machines --- a mechanism formally analogous to continuous normalizing flows \citep{https://doi.org/10.48550/arxiv.2302.00482} and known to reduce mixing times \citep{Ao_2004, Xing_2010} (\href{https://pni-lab.github.io/fep-attractor-network/appendix/#appendix-8}{Appendix~8}). The relationship to hierarchical predictive coding --- which minimizes the same objective in a bidirectional hierarchy rather than a fully recurrent topology --- is discussed in \href{https://pni-lab.github.io/fep-attractor-network/appendix/#appendix-8}{Appendix~8} and constitutes a promising direction for future work.
In general, there is growing interest in predictive coding-based neural network architectures - akin to the architecture presented herein. Recent studies have demonstrated that such approaches can not only reproduce backpropagation as an edge case \citep{https://doi.org/10.48550/arxiv.2202.09467}, but scale efficiently - even for cyclic graphs - and can outperform traditional backpropagation approaches in several scenarios \citep{https://doi.org/10.48550/arxiv.2308.07870, https://doi.org/10.48550/arxiv.2109.08063}. Furthermore, in our framework - in line with recent formulations of predictive coding for deep learning \citep{https://doi.org/10.48550/arxiv.2202.09467} - learning and inference are not disparate processes but rather two complementary facets of variational free energy minimization through active inference. As we demonstrated with simulations, this unification naturally endows the proposed architecture with characteristics of continual or lifelong learning - the ability for machines to gather data and fine-tune its internal representation continuously during functioning \citep{https://doi.org/10.48550/arxiv.2302.00487}.

A natural question is whether the structural properties of \acrshort{fep}-based self-orthogonalizing attractor networks confer robustness to adversarial perturbations, noise corruption, or data poisoning --- topics of growing importance in secure and trustworthy AI \citep{https://doi.org/10.48550/arxiv.1412.6572}. Within the Bayesian framing of the network (eq. (\ref{posterior-distribution})), adversarial perturbations and data poisoning correspond to attacks at two distinct levels: sensory perturbations (including adversarial examples) enter as erroneous bias shifts $\delta\mathbf{s}$ during inference (corrupted likelihood), whereas data poisoning distorts the learned prior $p(\boldsymbol{\sigma})$, i.e. the attractor landscape itself. The inverse-temperature parameter $iT$ mediates the balance between these levels: high precision deepens attractor basins relative to any bias perturbation, so the magnitude of $\delta\mathbf{s}$ becomes small compared to the basin depth; low precision, conversely, flattens the prior landscape and lets the actual sensory evidence dominate, mitigating the effect of distorted or poisoned attractors. This precision-mediated trade-off is a direct consequence of the Bayesian posterior structure and has no counterpart in deterministic Hopfield-type retrieval (see \href{https://pni-lab.github.io/fep-attractor-network/appendix/#appendix-7}{Appendix~7} for formal analysis).
Furthermore, our framework naturally distinguishes between two types of redundancy. Self-orthogonalization minimizes \textit{representational redundancy} (attractor overlap) which maximizes inter-attractor distances and thereby the ``adversarial budget'', i.e. the minimum perturbation energy needed to push the network across a basin boundary (\href{https://pni-lab.github.io/fep-attractor-network/appendix/#appendix-7}{Appendix~7}). On the other hand, the network maintains high \textit{structural redundancy}: the $M<N$ attractors are supported by $O(N^2)$ distributed weights, so bounded weight corruption or node failure has limited impact on the energy landscape.

Stochasticity - a key property of our network - is also very relevant from the perspective of artificial intelligence research. In our framework, noise is not an enemy; it implements the precision of inference, allowing it to strike a balance between stability and flexibility. This inherent stochasticity yields an exceptional fit with energy-efficient neuromorphic architectures \citep{Schuman_2022}, particularly within the emerging field of thermodynamic computing \citep{Melanson_2025} and memristive technologies, where attractor networks --- including memristive Hopfield networks, multi-wing and multi-butterfly hyperchaotic constructions, and memristive cellular-neural-network-like circuits \citep{Lin_2023, Wang_2023, Lin_2024, Deng_2025, Diao_2024} have recently been studied. A promising direction is to ask whether local \acrshort{vfe}-minimizing updates can be instantiated in thermodynamic and memristive substrates, combining the principled inferential interpretation of the \acrshort{fep} framework with the hardware efficiency and rich non-equilibrium dynamics in these emerging paradigms.

As inference in the proposed network involves spontaneously ``replaying'' its own attractors (or sequences thereof), even if no external input is introduced (zero bias), the proposed architecture may naturally overcome catastrophic forgetting, a highly problematic phenomenon in artificial neural networks where a model abruptly loses previously acquired knowledge when learning new information, especially during sequential training on multiple tasks. To examine how these properties of \acrshort{fep}-based attractor networks scale to more complex, diverse, and extended learning scenarios is a promising direction for further studies.

Finally, the recursive nature of our \acrshort{fep}-based formal framework provides a principled way to build hierarchical, multi-scale attractor networks which may be exploited for boosting the efficiency, robustness and explainability of large-scale AI systems. To realize its full potential in artificial intelligence research, the proposed architecture must be scalable to large real-world datasets. The per-step computational cost of the \acrshort{fep}-based attractor network is $O(N^2)$ for a full-network update --- dominated by one matrix-vector product for inference and one outer-product update for learning --- with no free-running phase required (unlike contrastive divergence in Boltzmann machines; see \href{https://pni-lab.github.io/fep-attractor-network/appendix/#appendix-8}{Appendix~8}). This complexity is confirmed empirically: runtime scales quadratically with $N$ across network sizes from 64 to 4096 nodes (\href{https://pni-lab.github.io/fep-attractor-network/simulation-scaling-jax}{Simulation 5}, \href{https://pni-lab.github.io/fep-attractor-network/appendix/#appendix-9}{Appendix~9}). Memory capacity benefits from emergent orthogonalization, which progressively approaches the projector-network limit of $K_{\max} = N$, substantially exceeding the classical Hopfield bound of $\sim 0.14N$ (\href{https://pni-lab.github.io/fep-attractor-network/simulation-scaling-jax}{Simulation 5}). We further validated the framework on the Olivetti faces dataset at full resolution ($N = 4096$, 400 patterns; \href{https://pni-lab.github.io/fep-attractor-network/simulation-faces-jax}{Simulation 6}), confirming that orthogonalization, Bayesian retrieval, and generalization transfer to more naturalistic stimuli.

In general, the proposed architecture inherently embodies all the previously discussed advantages of active inference and predictive coding \citep{https://doi.org/10.48550/arxiv.2202.09467, https://doi.org/10.48550/arxiv.2308.07870}. At every level of description the network dynamically balances accuracy and complexity and may naturally exhibit ``information-seeking'' behaviors (curiosity) \citep{Friston_2017}. Furthermore, the architecture may offer a foundation for exploring long-term philosophical implications of the qualities of active inference associated with sentience \citep{Pezzulo_2024}.

Despite the promise, several limitations in terms of scalability warrant explicit acknowledgment. First, the $O(N^2)$ memory footprint for the full weight matrix becomes prohibitive for very large $N$; sparse or factored weight parametrizations and multi-scale training approaches are a natural extension to boost scalability, but remain unexplored in the present work. Second, while solenoidal flows from asymmetric couplings are expected to accelerate mixing (\href{https://pni-lab.github.io/fep-attractor-network/appendix/#appendix-8}{Appendix~8}), a rigorous characterization of mixing times as a function of network size, pattern count, and coupling asymmetry is needed. Third, the current demonstrations use stationary or slowly varying input statistics; performance under strongly non-stationary, streaming real-world data --- where the input distribution shifts faster than the learning rate can track --- remains an open question. Fourth, all simulations use a single-layer architecture; scaling to hierarchical (multi-layer) deep particular partitions, which could support richer generative models and more efficient scaling, is an important theoretical and practical frontier. Finally, a key computational bottleneck is the sequential dependency of the Gibbs-like node-by-node inference and learning: each node's new state depends on the current states of all others, preventing straightforward parallelization. Our synchronous (full-network) JAX implementation sidesteps this by applying all updates simultaneously, yielding favorable empirical scaling up to $N = 4096$ (Simulation 5); however, this parallel schedule is an approximation whose fidelity at very large $N$ warrants further study. To efficiently scale the framework towards billions of neurons two complementary directions appear most promising: (i) dedicated hardware --- in particular thermodynamic computers \citep{Melanson_2025} and memristive substrates \citep{Lin_2023, Lin_2024} --- which can natively implement the local, stochastic update rules without the sequential bottleneck; and (ii) structured weight sparsity, which would reduce the per-step cost from $O(N^2)$ to $O(Ns)$ (where $s \ll N$ is the average number of non-zero connections per node) and simultaneously lower the memory footprint.

In conclusion, by deriving the emergence of adaptive, self-organizing - and self-orthogonalizing - attractor networks from the \acrshort{fep}, this work offers a principled synthesis of self-organization, Bayesian inference, and neural computation. The intrinsic tendency towards attractor orthogonalization, the multi-scale dynamics, and the continuous learning capabilities present a compelling, theoretically grounded outlook for better understanding natural intelligence and inspiring artificial counterparts through the lens of active inference.

\section{Manuscript Information}

\subsection{Data availability}

Simulation 1 and 5 is based on simulated data.
Simulation 2-4 is based on the `handwritten digits' dataset available in scikit-learn (\href{https://scikit-learn.org/stable/modules/generated/sklearn.datasets.load\_digits.html}{https://scikit-learn.org/stable/modules/generated/sklearn.datasets.load\_digits.html}), and originally published as part of the thesis of C. Kaynak., 1995: \href{https://archive.ics.uci.edu/dataset/80/optical+recognition+of+handwritten+digits}{https://archive.ics.uci.edu/dataset/80/optical+recognition+of+handwritten+digits}.
Simulation 6 is based on the Olivetti faces dataset, originally created at AT\&T Laboratories Cambridge and obtained via scikit-learn: \href{https://scikit-learn.org/0.19/datasets/olivetti\_faces.html}{https://scikit-learn.org/0.19/datasets/olivetti\_faces.html}.
The code for all simulations is available at \href{https://github.com/pni-lab/fep-attractor-network}{https://github.com/pni-lab/fep-attractor-network}. The repository contains a reference Python implementation of \acrshort{fep}-based self-orthogonalizing attractor networks (optimized for conceptual clarity); a separate scaling benchmark notebook uses a parallelized JAX update for runtime profiling only.

\subsubsection{Interactive manuscript}

\href{https://pni-lab.github.io/fep-attractor-network/}{https://pni-lab.github.io/fep-attractor-network/}

\subsubsection{Simulation source code}

\begin{itemize}
\item Simulation 1: \href{https://pni-lab.github.io/fep-attractor-network/simulation-demo}{https://pni-lab.github.io/fep-attractor-network/simulation-demo}
\item Simulation 2: \href{https://pni-lab.github.io/fep-attractor-network/simulation-digits}{https://pni-lab.github.io/fep-attractor-network/simulation-digits}
\item Simulation 3: \href{https://pni-lab.github.io/fep-attractor-network/simulation-digits-continuous-sequence}{https://pni-lab.github.io/fep-attractor-network/simulation-digits-continuous-sequence}
\item Simulation 4: \href{https://pni-lab.github.io/fep-attractor-network/simulation-digits-catastrophic-forgetting}{https://pni-lab.github.io/fep-attractor-network/simulation-digits-catastrophic-forgetting}
\item Simulation 5: \href{https://pni-lab.github.io/fep-attractor-network/simulation-scaling-jax}{https://pni-lab.github.io/fep-attractor-network/simulation-scaling-jax}
\item Simulation 6: \href{https://pni-lab.github.io/fep-attractor-network/simulation-faces-jax}{https://pni-lab.github.io/fep-attractor-network/simulation-faces-jax}
\end{itemize}
\printglossaries

\section*{Acknowledgments}
TS was supported by funding from the Deutsche Forschungsgemeinschaft (DFG, German Research Foundation) --- Project-ID 422744262 - TRR 289; and Project-ID 316803389 -- SFB 1280 ``Extinction Learning''. KF is supported by funding from the Wellcome Trust (Ref: 226793/Z/22/Z).

\bibliographystyle{unsrtnat}
\bibliography{main}

\newpage

\makeatletter
\let\@fnsymbol\@arabic
\makeatother

\author{}

\renewcommand{\headeright}{}
\renewcommand{\undertitle}{}
\renewcommand{\shorttitle}{}

\hypersetup{
pdftitle={\@title},
pdfsubject={},
pdfkeywords={attractor networks,free energy principle,active inference,orthogonal representations,self-organization},
addtopdfcreator={Written in Curvenote}
}

\setcounter{secnumdepth}{-1}

\maketitle
\footnotetext[1]{Correspondence to: tamas.spisak@uk-essen.de}

\section{Appendix}

Library imports for inline code.

\begin{verbatim}
import numpy as np
import seaborn as sns
from matplotlib import pyplot as plt
import scipy.integrate as integrate

sns.set(style="white")
\end{verbatim}

\section{Appendix 1}

\textbf{Continuous Bernoulli distribution}
\newline
\newline
$x \sim \mathcal{CB}(L) \iff P(x) \propto e^{Lx}, \quad x \in [ -1, 1]\subset \mathbb{R}$
\newline
\newline
Below we provide a Python implementation of the continuous Bernoulli distributions: $\mathcal{CB}(L)$, parametrized by the log odds L and adjusted to the [-1,1] interval.

\begin{verbatim}
def CB(x, b, logodds=True):
    if not logodds:
        b=np.log(b/(1 -b))
    if np.isclose(b, 0):
        return np.ones_like(x)/2
    else:
        return b * np.exp(b*x) / (2*np.sinh(b)) 
    
# Plot some examples
delta = 100
sigma = np.linspace( -1, 1, delta)
for b in np.linspace( -2, 2, 5):
    p_sigma = CB(sigma, b)
    sns.lineplot(x=sigma, y=p_sigma, label=b)
\end{verbatim}

\includegraphics[width=0.7\linewidth]{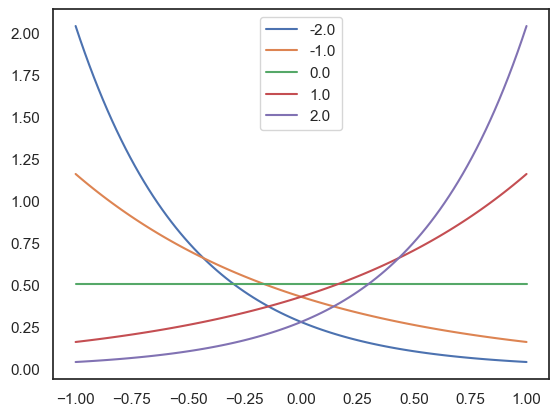}

\section{Appendix 2}

\textbf{Continuous Bernoulli distribution full derivation}

Derivation of the exponential form of the continuous Bernoulli distribution, parametrized by the log odds L and adjusted to the [-1,1] interval.

$P(x; L) = \frac{e^{Lx}}{\int_{ -1}^1 e^{Lx}} = \frac{e^{Lx}}{ \frac{e^L - e^{ -L}}{L}} = L \frac{e^{Lx}}{2sinh(L)}$

\section{Appendix 3}

\textbf{Visualization of the likelihood function with various parameters, in python.}

\begin{verbatim}
delta = 100
s_i = np.linspace( -1, 1, delta)

fig, axes = plt.subplots(1, 3, figsize=(12, 3), sharey=True)
for idx, mu in enumerate([0.1, 0.5, 1]):
    for w_i in np.linspace( -2, 2, 5):
            p_mu = CB(s_i, w_i*mu)
            sns.lineplot(x=s_i, y=p_mu, ax=axes[idx], label=w_i).set(title=f"$\\mu={mu}$")
\end{verbatim}

\includegraphics[width=0.7\linewidth]{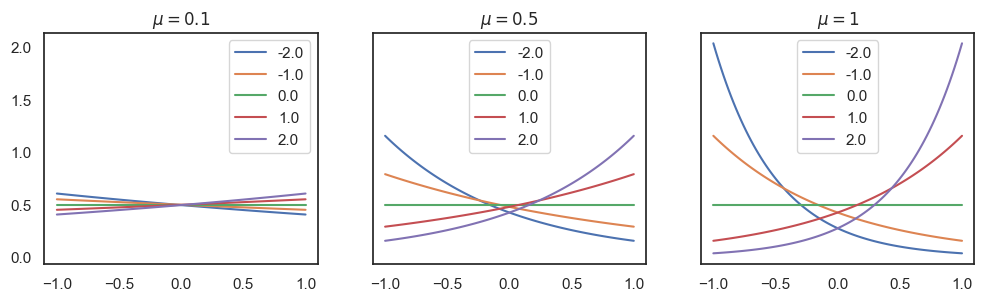}

\section{Appendix 4}

\textbf{Expected value of the $\mathcal{CB}$}

\begin{align}
\mathbb{E}_{\mathcal{CB}(b)}[\sigma] &=  \int \sigma \frac{ e^{b \sigma} }{ 2sinh(b) } d \sigma
\\ &= \frac{ b \left( \frac{(b-1)e^b}{b^2} + \frac{(b+1)e^-b}{b^2} \right) }{ 2sinh(b) } \\
&= \coth(b) - \frac{1}{b}
\end{align}

\begin{verbatim}
bs = np.linspace( -10, 10, 100)
plt.figure(figsize=(4, 1))
sns.lineplot(x=bs, y=1/np.tanh(bs) - 1/bs) # coth(b) - 1/b == 1/2 sech^2(b)
plt.show()
\end{verbatim}

\includegraphics[width=0.7\linewidth]{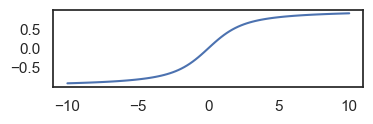}

\section{Appendix 5}

\textbf{Conservative dynamics}

Let $x \in \mathbb{R}^n$ denote the system's states (internal, blanket, and external states).
The drift can be decomposed into a gradient part, from a potential $U$, and a solenoidal part $R$:
\[
\dot{x} = -\nabla U(x) + R(x).
\]
Here $R = -R^{\mathsf{T}}$ is antisymmetric in state space.

The probability density $p(x,t)$ over states evolves according to the Fokker--Planck equation:
\[
\partial_t p(x,t)
= -\nabla \cdot \left[ \left(-\nabla U(x) + R(x)\right)p(x,t) \right]
+ \text{diffusion terms}.
\]

To determine the stationary distribution $p_s(x)$, we set $\partial_t p(x,t)=0$.
The Fokker--Planck equation then implies that the net probability current
\[
J_s(x) = \left(-\nabla U(x) + R(x)\right)p_s(x) - D\nabla p_s(x)
\]
assuming isotropic diffusion $D$, must be divergence-free:
\[
\nabla \cdot J_s(x) = 0.
\]
If we propose $p_s(x)=Z^{-1}\exp(-U(x)/D)$, where $Z$ is a normalization constant, then the diffusive current $-D\nabla p_s(x)$ becomes $p_s(x)\nabla U(x)$.

Substituting this into $J_s(x)$, we get:
\[
J_s(x)
= \left(-\nabla U(x) + R(x)\right)p_s(x) + p_s(x)\nabla U(x)
= R(x)p_s(x).
\]

Thus, $p_s(x) \propto \exp(-U(x)/D)$ is the stationary distribution if and only if the solenoidal component of the probability current, $J_R(x)=R(x)p_s(x)$, is itself divergence-free:
\[
\nabla \cdot \left(R(x)p_s(x)\right) = 0.
\]

It is a key result from the study of non-equilibrium systems under the Free Energy Principle, particularly those possessing a Markov blanket and involving conservative particles, that this condition $\nabla \cdot (R(x)p_s(x)) = 0$ holds (see \citep{Friston_2023}; but also related: \citet{Ao_2004}; \citet{Xing_2010}). The conditional independence structure imposed by the Markov blanket, along with other FEP-specific assumptions, such as conservative particles, constrains the system's dynamics such that solenoidal forces $R(x)$ do not alter the Boltzmann form.

In most NESS systems, antisymmetric flows do alter the stationary measure. Under particular‐partition (Markov‐blanket) constraints, however, the internal--external factorization ensures that solenoidal (antisymmetric) flows remain divergence‐free, leaving the Boltzmann‐like steady distribution intact.
This is why, in these Markov‐blanketed systems, one can have persistent solenoidal currents (nonequilibrium flows) yet preserve a stationary distribution that depends only on the symmetric part of the couplings.

\section{Appendix 6}

\textbf{Detailed inference derivation: accuracy--complexity decomposition and $\dfrac{\partial F}{\partial b_q}$}
\newline
\newline
The main text derives the inference rule (eq. (\ref{inference-rule})) via a compact local VFE argument. Here we provide the equivalent --- but more detailed --- derivation via the accuracy--complexity decomposition of VFE, confirming the same result.
\newline
\newline
\textbf{1. Subsitute our parametrization into F}
\newline
\newline
Let's start with substituting our parametrization into eq. \href{\#f-complexity-accuracy}{}.
\newline
\newline
\textbf{1a. Accuracy term}
\newline
\newline
From the RBM marginalization (eq. \href{\#rmb-to-hopfield}{}):
\newline
\newline
$E(\bm{\sigma}) = \underbrace{ -b_i\sigma_i - \sum_{j\neq i}J_{ij}\sigma_i\sigma_j}_{\text{Terms with } \sigma_i} \underbrace{ -\sum_{j\neq i}b_j\sigma_j - \frac{1}{2}\sum_{j,k\neq i}J_{jk}\sigma_j\sigma_k}_{\text{Terms without } \sigma_i}$
\newline
\newline
Here, $-b_i\sigma_i$ becomes constant, since $\sigma_i$ is fixed. So we get:
\newline
\newline
$P(\sigma_{\backslash i}|\sigma_i) \propto \exp\left(\sum_{j\neq i}(b_j + J_{ij}\sigma_i)\sigma_j + \frac{1}{2}\sum_{j,k\neq i}J_{jk}\sigma_j\sigma_k\right)$
\newline
\newline
Taking expectation of $\ln P(\sigma_{\backslash i}|\sigma_i)$ under $q( \sigma_i)$:
\newline
\newline
$\mathbb{E}q[\ln P( \sigma_{\backslash i}| \sigma_i)] = \text{const} + \sum_{j\neq i}b_j \sigma_j + S(b_q)\sum_{j\neq i}J_{ij} \sigma_j + \frac{1}{2}\sum_{j,k\neq i}J_{jk} \sigma_j \sigma_k$
\newline
\newline
Where $S(b_q) = \mathbb{E}_q[ \sigma_i] = \coth b_q - 1/b_q$ is the expected value of the $\mathcal{CB}$, a sigmoid function of the bias (\#supplementary-information-4)).
\newline
\newline
\textbf{1b. Complexity term}
\newline
\newline
The complexity term in eq. \href{\#f-complexity-accuracy}{} is simply the KL-divergence term between two $\mathcal{CB}$ distributions. For $\mathcal{CB}$ distributions:
\newline
\newline
$q(x) = \frac{b_q}{2\sinh b_q}e^{b_q x},\quad p(x) = \frac{b}{2\sinh b}e^{b x}$
\newline
\newline
$\cdot$ KL divergence definition:
\newline
\newline
$D_{KL} = \int_{ -1}^1 q(x) \ln\frac{q(x)}{p(x)} dx = \mathbb{E}_q[\ln q(x) - \ln p(x)]$
\newline
\newline
$\cdot$ Expand log terms:
\newline
\newline
$\ln q(x) = \ln b_q - \ln(2\sinh b_q) + b_q x$
\newline
\newline
$\ln p(x) = \ln b - \ln(2\sinh b) + b x$
\newline
\newline
$\cdot$ Subtract log terms:
\newline
\newline
$\ln\frac{q(x)}{p(x)} = \ln\frac{b_q}{b} + \ln\frac{\sinh b}{\sinh b_q} + (b_q - b)x$
\newline
\newline
$\cdot$ Take expectation under q(x):
\newline
\newline
$D_{KL} = \ln\frac{b_q}{b} + \ln\frac{\sinh b}{\sinh b_q} + (b_q - b)\mathbb{E}_q[x]$
\newline
\newline
$\cdot$ Compute expectation $\mathbb{E}_q[x]$:
\newline
\newline
$\mathbb{E}_q[x] = \int_{ -1}^1 x \frac{b_q e^{b_q x}}{2\sinh b_q} dx = \frac{1}{2\sinh b_q}\left[\frac{e^{b_q x}}{b_q^2}(b_q x - 1)\right]_{ -1}^1$
\newline
\newline
$\cdot$ Evaluate at bounds:
\newline
\newline
$= \frac{1}{2\sinh b_q}\left(\frac{e^{b_q}(b_q - 1) - e^{ -b_q}( -b_q - 1)}{b_q^2}\right)$
\newline
\newline
$\cdot$ Simplify using hyperbolic identities:
\newline
\newline
$= \frac{(b_q \cosh b_q - \sinh b_q)}{b_q^2 \sinh b_q} = \coth b_q - \frac{1}{b_q}$
\newline
\newline
$\cdot$ Final substitution for the complexity term:
\newline
\newline
$D_{KL} = \ln\frac{b_q \sinh b}{b \sinh b_q} + (b_q - b)\left(\coth b_q - \frac{1}{b_q}\right)$
\newline
\newline
\textbf{1c. Combining the two terms}
\newline
\newline
Combining the two terms, we get the following expression for the free energy:
\newline
\newline
$F = \ln\left(\frac{b_q}{b}\right) + \ln\left(\frac{\sinh(b)}{\sinh(b_q)}\right) + (b_q - b) S(b_q) - \sum_{j \ne i} \left( b_j + S(b_q) J_{ij} \right) \sigma_j - \dfrac{1}{2} \sum_{j \ne i} \sum_{k \ne i} J_{jk} \sigma_j \sigma_k + C$
\newline
\newline
where C denotes all constants in the equation that are independent of $\sigma or b_q$.
\newline
\newline
\textbf{2. Free Energy partial derivative calculation}
\newline
\newline
$\cdot$ First, we differentiate the log terms:
\newline
\newline
$\frac{\partial}{\partial b_q}\left[\ln\frac{b_q}{b} + \ln\frac{\sinh b}{\sinh b_q}\right] = \frac{1}{b_q} - \coth b_q$
\newline
\newline
$\cdot$ Then, the KL core term:
\newline
\newline
$\frac{\partial}{\partial b_q}\left[(b_q - b)S(b_q)\right] = S(b_q) + (b_q - b)\frac{dS}{db_q}$
\newline
\newline
$\cdot$ The linear terms:
\newline
\newline
$\frac{\partial}{\partial b_q}\left[ -\sum_{j \neq i}(b_j + S(b_q)J_{ij})\sigma_j\right] = -\sum_{j \neq i}J_{ij}\sigma_j\frac{dS}{db_q}$
\newline
\newline
$\cdot$ The constants vanish.
\newline
\newline
$\cdot$ Now, combining all terms:
\newline
\newline
$\frac{\partial F}{\partial b_q} = \left(\frac{1}{b_q} - \coth b_q\right) + \left(S(b_q) + (b_q - b)\frac{dS}{db_q}\right) - \sum_{j \neq i}J_{ij}\sigma_j\frac{dS}{db_q}$
\newline
\newline
$\cdot$ Substituting $S(b_q) = \coth b_q - 1/b_q$:
\newline
\newline
$= \left(\frac{1}{b_q} - \coth b_q\right) + \left(\coth b_q - \frac{1}{b_q} + (b_q - b)\frac{dS}{db_q}\right) - \sum_{j \neq i}J_{ij}\sigma_j\frac{dS}{db_q}$
\newline
\newline
$\cdot$ Cancel terms:
\newline
\newline
$\cancel{\frac{1}{b_q}} \cancel{ - \coth b_q} + \cancel{\coth b_q} \cancel{ - \frac{1}{b_q}} + (b_q - b)\frac{dS}{db_q} - \sum_{j \neq i}J_{ij}\sigma_j\frac{dS}{db_q}$
\newline
\newline
Gives us the \textbf{final derivative}:
\newline
\newline
$\frac{ \partial F}{ \partial b_q} = \left(b_q - b - \sum_{j\neq i}J_{ij} \sigma_j\right)\frac{dS}{db_q}$
\newline
\newline
Where $\frac{dS}{db_q} = -csch^2 b_q + \frac{1}{b_q^2}$.
\newline
\newline
Setting the derivative to zero and solving for $b_q$, we get:
$b_q = b + \sum_{j \ne i} J_{ij} \sigma_j$
\newline
\newline
Now we remember that the expected value of the $\mathcal{CB}$ is the Langevin function of its bias -, $\mathbb{E}(x) = coth(b) - 1/b$ (\href{\#supplementary-information-4}{}). For simplicity, we will denote it as $L(x)$. Now we can write:
\newline
$\mathbb{E}_{q}[\sigma_i] = L(b_q) = L \left( b + \sum_{j \ne i} J_{ij} \sigma_j \right)$
\newline
\newline
\textbf{Q.E.D.}

\section{Appendix 7}

\textbf{Basin geometry and adversarial robustness}
\newline
\newline
We provide a geometric analysis of how attractor orthogonality and the precision parameter jointly affect adversarial robustness in the proposed framework.
\newline
\newline
\textbf{Setup.} Consider a network with $N$ nodes, inverse temperature $iT$, and $K$ stored attractors $\{\boldsymbol{\sigma}^{(\mu)}\}_{\mu=1}^K$ with $\|\boldsymbol{\sigma}^{(\mu)}\|^2 = N$. Under the projector weight matrix $\mathbf{J}^\dagger \approx \frac{1}{N}\sum_\mu \boldsymbol{\sigma}^{(\mu)}{\boldsymbol{\sigma}^{(\mu)}}^\top$ (zero diagonal), which the learning rule approximates, the energy at a stored attractor satisfies $E(\boldsymbol{\sigma}^{(\mu)}) \approx -N$.
\newline
\newline
\textbf{Adversarial budget and representational redundancy.} An adversarial perturbation $\delta\mathbf{s}$ applied to the bias vector shifts the energy difference between two attractors $\boldsymbol{\sigma}^{(\mu)}$ and $\boldsymbol{\sigma}^{(\nu)}$ by $\Delta E \approx -\delta\mathbf{s}^\top (\boldsymbol{\sigma}^{(\mu)} - \boldsymbol{\sigma}^{(\nu)})$. The most efficient attack aligns $\delta\mathbf{s}$ with the difference vector, yielding a minimum perturbation norm $\|\delta\mathbf{s}\|_{\min} \propto 1/\|\boldsymbol{\sigma}^{(\mu)} - \boldsymbol{\sigma}^{(\nu)}\|$. For orthogonal attractors ($\boldsymbol{\sigma}^{(\mu)\top}\boldsymbol{\sigma}^{(\nu)} = 0$): $\|\boldsymbol{\sigma}^{(\mu)} - \boldsymbol{\sigma}^{(\nu)}\|^2 = 2N$. For attractors with normalized overlap $m = \frac{1}{N}\boldsymbol{\sigma}^{(\mu)\top}\boldsymbol{\sigma}^{(\nu)}$: $\|\boldsymbol{\sigma}^{(\mu)} - \boldsymbol{\sigma}^{(\nu)}\|^2 = 2N(1 -m)$. Since $0 \leq m < 1$, orthogonal attractors ($m = 0$) maximize the adversarial budget. Self-orthogonalization thus reduces \textit{representational} redundancy (overlap between attractor states) while \textit{increasing} robustness to targeted perturbations.
\newline
\newline
\textbf{Structural redundancy.} Although representational redundancy is minimized, \textit{structural} redundancy is fully preserved: each attractor is encoded across all $O(N^2)$ synaptic weights, with each weight contributing $O(1/N)$ to the basin depth. Corrupting or removing a bounded fraction of weights therefore has a limited effect on the energy landscape, regardless of attractor orthogonality.
\newline
\newline
\textbf{Prior precision as a defense.} Under finite temperature, the posterior is $p(\boldsymbol{\sigma}\mid\mathbf{s}) \propto \exp\bigl\{iT\bigl[\sum_i (b_i + \delta s_i)\sigma_i + \frac{1}{2}\sum_{ij}J_{ij}\sigma_i\sigma_j\bigr]\bigr\}$. The effective basin depth scales as $iT \cdot N$, while the perturbation contributes $iT \cdot \delta\mathbf{s}^\top\boldsymbol{\sigma}$. The ratio of basin depth to perturbation magnitude grows with $iT$: high precision makes prior-encoded attractors robust to sensory-level attacks. The converse also holds: low $iT$ down-weights the prior, limiting the influence of poisoned attractors on the posterior.
\newline
\newline
\textbf{Stochastic averaging.} Beyond the geometric analysis, stochastic MCMC inference provides an additional defense layer. The time-averaged network response converges to the posterior mean $\mathbb{E}_{p(\boldsymbol{\sigma}\mid\mathbf{s})}[\boldsymbol{\sigma}]$, with variance decreasing as $O(1/T_{\text{eff}})$ where $T_{\text{eff}}$ is the effective number of independent samples. This averaging smooths out the effect of perturbations that shift individual samples but do not move the posterior mode across a basin boundary.
\newline
\newline
\textbf{Implicit regularization against data poisoning.} During learning, the anti-Hebbian term in the learning rule subtracts the network's current prediction from the observed correlation. For a corrupted training pattern, the prediction --- dominated by the clean attractor subspace --- absorbs most of the pattern's variance, leaving only a small residual update. For a poisoning fraction $\epsilon \ll 1$, the cumulative effect of poisoned updates remains bounded relative to the clean learning signal, analogous to the outlier-robustness of Bayesian posteriors under informative priors.

\section{Appendix 8}

\textbf{Detailed comparison with classical models}
\newline
\newline
We compare the proposed FEP-based attractor network (FEP-ANN) with the canonical formulations of classical Hopfield networks, Boltzmann machines, projector (pseudo-inverse) networks, and hierarchical predictive coding. Each of these model families encompasses many variants; the comparison targets the canonical baseline in each case.
\newline
\newline
\textbf{Learning rule and computational complexity}

\begin{itemize}
\item \textbf{Boltzmann machines (contrastive divergence).}
The exact gradient of the Boltzmann log-likelihood with respect to weights $J_{ij}$ is

\begin{equation}
\frac{\partial \ln p(\mathbf{v})}{\partial J_{ij}} = \langle \sigma_i \sigma_j \rangle_{\text{data}} - \langle \sigma_i \sigma_j \rangle_{\text{model}}
\end{equation}

where $\langle \cdot \rangle_{\text{model}}$ requires sampling from the model's equilibrium distribution --- computationally intractable for large networks. Contrastive divergence (CD-$k$) approximates $\langle \cdot \rangle_{\text{model}}$ by running $k$ steps of Gibbs sampling from the data-initialized state. Each Gibbs sweep visits all $N$ nodes, and computing the input to each node requires $O(N)$ multiplications, yielding a per-step cost of $O(N^2)$ per Gibbs sweep and $O(N^2 k)$ per weight update.

\item \textbf{FEP-ANN.} The learning rule for node $i$ uses the network's \textit{instantaneous} predicted state $\hat{\sigma}_i = L(b_i + \sum_k J_{ik} \sigma_k)$ --- a single forward pass through the current weights --- rather than a sample average over model-generated states. Computing $\hat{\sigma}_i$ for all nodes requires one matrix-vector product $\mathbf{J}\boldsymbol{\sigma}$ at $O(N^2)$; the outer-product updates for all weights then cost $O(N^2)$. The total per-step learning cost is therefore $O(N^2)$ --- a factor of $k$ cheaper than CD-$k$. The free-running phase is eliminated entirely.
\item \textbf{Solenoidal flows and mixing times}
In a symmetric Boltzmann machine, inference proceeds via reversible Gibbs sampling; reversibility implies that probability mass can only move between attractor basins by climbing over energy barriers, leading to mixing times that scale exponentially with barrier height.
\newline
\newline
In the FEP-ANN with asymmetric couplings $\mathbf{J} \neq \mathbf{J}^\top$, the dynamics decompose into a gradient (dissipative) component driven by $\mathbf{J}^\dagger = \frac{1}{2}(\mathbf{J}+\mathbf{J}^\top)$ and a solenoidal (conservative) component driven by $\mathbf{J}^ - = \frac{1}{2}(\mathbf{J} -\mathbf{J}^\top)$. The solenoidal term does not alter the stationary distribution but induces persistent probability currents along iso-energy contours. These non-reversible currents transport probability mass between attractor basins \textit{without} climbing energy barriers --- they traverse along the iso-energy surface instead, accelerating mixing. The resulting dynamics are formally analogous to continuous normalizing flows and Markovian flow matching.
\item \textbf{Orthogonalization and memory capacity}
Classical Hopfield networks with Hebbian (outer-product) weights store $K$ patterns; cross-talk limits capacity to $K_{\max} \approx 0.14N$. Projector (pseudo-inverse) networks achieve optimal capacity $K_{\max} = N$ with error-free retrieval but require batch access and $O(N^3)$ computation. In the FEP-ANN, the Hebbian/anti-Hebbian learning rule incrementally approximates the projector matrix through online learning: the Hebbian term accumulates pattern correlations while the anti-Hebbian term subtracts the model's current prediction, progressively decorrelating the stored representations. Memory capacity thus approaches the projector limit without batch computation or matrix inversion.
\item \textbf{Relationship to hierarchical predictive coding}
Both the FEP-ANN and hierarchical predictive coding (PC) networks minimize variational free energy via local, prediction-error-driven updates. The structural difference is \textit{topology}: PC operates on a bidirectional hierarchy with directed inter-layer connections (feedforward predictions, feedback prediction errors), while the FEP-ANN uses a fully recurrent, single-layer topology with arbitrary lateral couplings. Within the deep particular partition formalism, both are instances of the same FEP-based inference architecture applied to different coupling topologies. Exploring the formal equivalences between these topologies is a promising direction for future work.
\end{itemize}

\begin{table}\scriptsize
\centering
\caption[]{Comparison of the FEP-ANN with canonical formulations of classical models.}
\label{tab-comparison}
\begin{tabular}{p{\dimexpr 0.167\linewidth-2\tabcolsep}p{\dimexpr 0.167\linewidth-2\tabcolsep}p{\dimexpr 0.167\linewidth-2\tabcolsep}p{\dimexpr 0.167\linewidth-2\tabcolsep}p{\dimexpr 0.167\linewidth-2\tabcolsep}p{\dimexpr 0.167\linewidth-2\tabcolsep}}
\toprule
Feature & Hopfield & Boltzmann machine & Projector network & Predictive coding & \textbf{FEP-ANN (ours)} \\
\hline
\textbf{Derivation} & Energy-based & Statistical mechanics & Optimal storage & Hierarchical Bayesian & First principles (FEP) \\
\textbf{State space} & Binary $\{ -1,+1\}$ & Binary $\{0,1\}$ & Binary & Continuous (Gaussian) & Continuous $[ -1,+1]$ (CB) \\
\textbf{Activation} & Sign & Logistic sigmoid & Sign & Linear / nonlinear & Langevin (emergent) \\
\textbf{Learning} & Hebbian (batch) & Contrastive divergence & Pseudo-inverse (batch) & Prediction-error min. & Hebbian/anti-Hebbian (online) \\
\textbf{Learning phases} & One-shot & Clamped + free & One-shot & Layerwise & Single phase \\
\textbf{Coupling} & Symmetric & Symmetric & Symmetric & Directed (hierarchy) & Symmetric or asymmetric \\
\textbf{Sequence dynamics} & No & No & No & Temporal hierarchy & NESS solenoidal flow \\
\textbf{Orthogonalization} & No & No & By construction & No & Emergent (VFE) \\
\textbf{Memory capacity} & $\sim 0.14N$ & $\sim 0.14N$ & $N$ (optimal) & N/A (generative) & Approaches $N$ \\
\textbf{Inference} & Deterministic & Gibbs MCMC & Deterministic & Message passing & MCMC + solenoidal flow \\
\textbf{Precision control} & None & Temperature (fixed) & None & Precision-weighting & $iT$ (Bayesian) \\
\textbf{Continual learning} & Catastrophic forgetting & Catastrophic forgetting & Catastrophic forgetting & Possible (with replay) & Built-in (spontaneous replay) \\
\textbf{Learning complexity} & $O(KN^2)$ one-shot & $O(N^2 k)$ per step & $O(KN^2 + N^3)$ one-shot & $O(N_l)$ per layer per step & $O(N^2)$ per step \\
\textbf{FEP / active inference} & No formal link & Variational (post hoc) & No formal link & Compatible & Derived from FEP \\
\bottomrule
\end{tabular}
\end{table}

\section{Appendix 9}

\textbf{Scaling benchmarks}
\newline
\newline
We report runtime and memory capacity benchmarks for the JAX implementation across network sizes from $N = 64$ to $N = 4096$ and pattern counts from $K = 10$ to $K = N$.
Training and inference wall-clock time as a function of network size $N$, confirming the expected $O(N^2)$ per-step scaling. \textit{Panel to be finalized with benchmark data from} 07-simulation-scaling-jax.ipynb.
\newline
\newline
\textbf{Table: Runtime benchmarks (JAX implementation, 10 updates)}

\bigskip\noindent
\begin{tabular}{p{\dimexpr 0.333\linewidth-2\tabcolsep}p{\dimexpr 0.333\linewidth-2\tabcolsep}p{\dimexpr 0.333\linewidth-2\tabcolsep}}
\toprule
$N$ (nodes) & Training time (s) & Throughput (steps/s) \\
\hline
100 & 0.004 & 2455 \\
1 000 & 0.007 & 1390 \\
5 000 & 0.26 & 39.1 \\
10 000 & 0.39 & 25.5 \\
20 000 & 1.49 & 6.69 \\
50 000 & 59.8 & 0.167 \\
\bottomrule
\end{tabular}

\bigskip


\end{document}